\documentclass[]{aa}

\usepackage{amssymb}
\usepackage{graphicx}
\usepackage{longtable,lscape,supertabular}
\usepackage{txfonts}
\usepackage{natbib}

\begin{document}
\newcommand{\ergcms}{erg cm$^{-2}$ s$^{-1}$}
\newcommand{\ergs}{erg s$^{-1}$}

\title{Hot gas in groups: NGC 5328 and the intriguing case of NGC 4756 with XMM-Newton
\thanks{Based on
XMM-{\it Newton} observations  (Obs. ID 0551600101 AND 0401480201  P.I. G. Trinchieri)}}
\author{G. Trinchieri\inst{1}, A. Marino\inst{2}, P. Mazzei\inst{3},  
R. Rampazzo\inst{3}, 
A. Wolter\inst{1}} 
\institute{INAF-Osservatorio Astronomico di Brera, Via Brera 28, 20121 Milano, Italy\\
 \email{ginevra.trinchieri@brera.inaf.it; anna.wolter@brera.inaf.it}
\and 
Dipartimento di Fisica e Astronomia G. Galilei, Universit\`a  di Padova, vicolo dell'Osservatorio 3, I35122 Padova 
\email{antonietta.marino@oapd.inaf.it}
              \and
INAF-Osservatorio Astronomico di Padova, Vicolo dell'Osservatorio 5, 35122 Padova, Italy\\
 \email{paola.mazzei@oapd.inaf.it; roberto.rampazzo@oapd.inaf.it}
}
\date{Received ; accepted}
\authorrunning{Trinchieri et al.}
\titlerunning{NGC 5328 and NGC 4756 with XMM-Newton}

\abstract
{The environment appears to have a strong influence on fundamental
properties of galaxies, modifying their morphology  and their  star
formation histories.  Similarly, galaxies play a role in determining
the properties of the hot intergalactic medium in groups, heating  and
enriching it through a variety of mechanisms. NGC 5238 and NGC 4756 are the brightest
unperturbed elliptical galaxies in their respective loose groups,
but analysis of their environment suggest that they might be at different evolutionary stages.}
{In the present study we aim at characterizing the properties of  the hot gas
in the halos of the brightest members and in the environment.  In NGC 4756  we are
also interested in the properties of a substructure identified to the SW and the region
connecting the two structures, to search for a physical connection between
the two.  However, we have to take into account the fact that
the group is projected against the bright, X-ray emitting cluster A1361,
which heavily contaminates and confuses the emission from the foreground
structure.}
  {We present XMM-{\it Newton} observations
of the groups and the careful analysis to separate different components.
We examine the X-ray morphology, hot gas distribution and  spectral
characteristics of NGC 4756 and NGC 5328 and their companion galaxies. To better characterize the ambient, we also present a re-evaluation of the dynamical properties of the systems. SPH simulations are used to interpret the results.}
  {We find that the X-ray source
associated with NGC 4756 indeed sits on top of extended emission from
the background cluster A1361, but can be relatively well distinguished
from it as a significant excess over it  out to $r \rm \sim150''\ (\sim 40\ kpc)$.    NGC~4756
has an X-ray luminosity of  $\rm  L_x\sim 10^{41}$ erg s$^{-1}$  due to hot gas, with an average
temperature of kT$\sim 0.7 $ keV. We measure a faint diffuse emission also in the region of the subclump to the SW, but more interestingly, we detect gas between the two structures, indicating a possible physical connection. The X-ray emission from NGC 5328 is clearly peaked on the galaxy, also at $\rm  L_x\sim 10^{41}$ erg s$^{-1}$, and extends to r $\sim 110$ kpc. 
Simulations provide an excellent reproduction of the SED and the global properties of both galaxies, which are caught at two different epochs of the same evolutionary process, with NGC 5328 $\sim 2.5$ Gyr younger than NGC 4756. }
{}
\maketitle

\section{Introduction}

Observationally, we see that a large fraction ($\sim$50-60 \%) of the galaxies 
in the Local Universe is in groups.  These could be isolated systems, or could be part of  filaments or chains, and  near/falling into clusters \citep[see e.g.][]{Eke04, Tago08}.  
Groups, for which the galaxy velocity dispersion is comparable to the internal  velocities in galaxies,  provide a  controlled environment in which interactions (mergers)  act to modify galaxy properties, for example by removing the gas that fuels star formation.  Recent surveys have indeed verified that the 
transformation from star-forming systems into  ``passive'' ones  \citep[][and references therein]{bai} is a result of the environmental 
action over the past 8 billion years, since redshift $z\sim1$ 
\citep{Kovac10,Peng10}. 
Cosmological hydrodynamical simulations by e.g. \cite{Kobayashi05} and \cite{Feldmann10, Feldmann11} to study the evolution of groups suggest that the  star-forming massive galaxies at the  center of group-type potentials at z$>>1 $ can become the massive, gas-poor early-type systems observed at the center of groups today.

   \begin{figure*}
 \includegraphics[width=\hsize,clip=true]{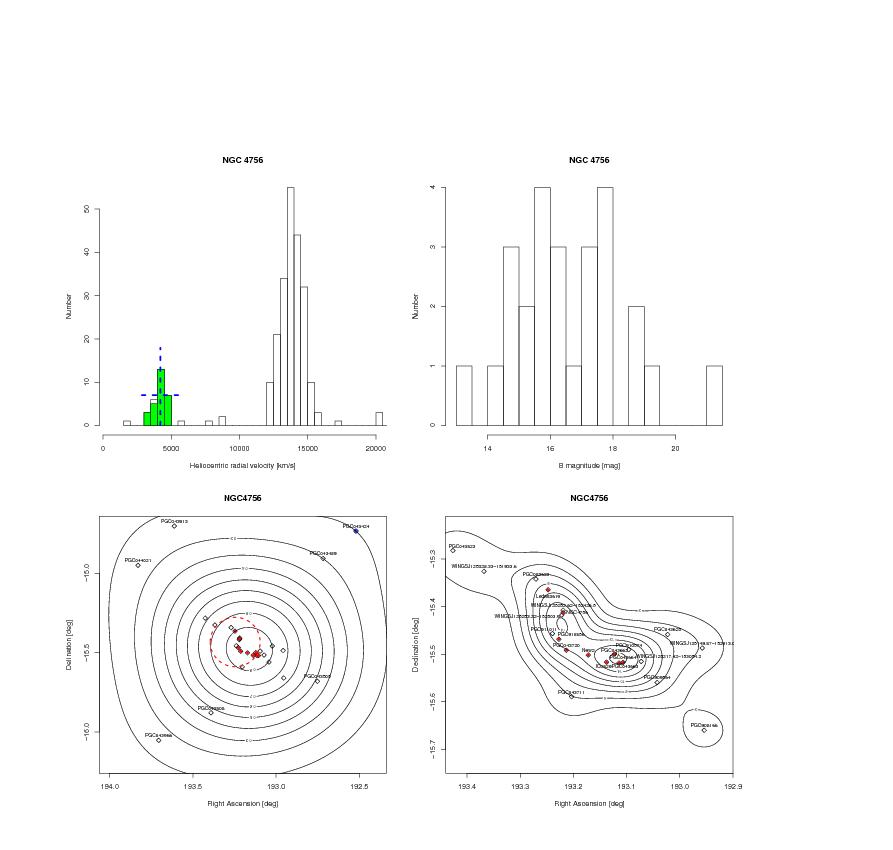}
      \caption{
Top Left: Heliocentric radial velocities in a 1.5 Mpc box centered on NGC 4756. The mean heliocentric radial velocity and the 3$\sigma$ velocity
dispersion of the group are identified (dashed lines). The second  peak 
is due to 
A1631. Top right: B magnitude distribution
of the group members.
Bottom panels:  spatial distribution of galaxies centered on NGC 4756 within
1.5 Mpc, and a zoom corresponding to the  XMM-{\it Newton} field of view (right),
superposed on the 2D binned kernel-smoothed number density contours.
Filled red squares are group members listed in \cite{GrutzbaN4756}. 
The dashed circle is centered on the center of mass of the group and identifies the
virial radius.
}
         \label{1pdf}
   \end{figure*}
%

We therefore can expect that galaxies that we observe in clusters today
are likely to have experienced  pre-processing
in groups at some point in their history, before the group  itself fell into the cluster formed within the same 
infalling halo. In their study of close pairs in 
the Sloan Digital Sky Survey (SDSS), \cite{Perez09} found  that galaxies are efficiently pre-processed by close
encounters and mergers while in intermediate-density
environments.
With the same mechanism of interaction between pairs,
mergers have probably depleted the reservoir of
galaxies in the halo while building the dominant elliptical of the
group.
In addition to mergers, simulations suggest that
ram-pressure stripping, once thought to
operate only in rich environments, is active in the form of  ``strangulation'' 
in groups as well.
This depletes gas rich galaxies from their hot ISM content, 
which is their largest gas 
reservoir, leaving  their molecular gas content nearly intact.
In a time scale of about
1 Gyr, this process leads to quenching of star formation in gas rich disk galaxies, transforming
them into S0s
 \citep{Kawata08}.
In fact  \citet{Jeltema08} interpret the evidence that the $L_K - L_X$ relation for early-type galaxies in groups is systematically below that of field objects 
as an indirect evidence of hot gas stripped by viscous- or ram-pressure. This is supported by  a more direct evidence of the action of a stripping event seen in the X-ray image of the S0 galaxy NGC 6265, located $\sim 250$ kpc west of the NGC6269 group (see also \citealt{baldi}, or \citealt{kim} for NGC 7619).

However, a different scenario is proposed by \cite{Berrier}.  Their  simulations actually suggest that only a 
small fraction (30\%) of galaxies today in cluster may have been accreted 
from groups, while the majority are accreted directly from the field and then reprocessed in the cluster environment.

To better understand processes related to galaxy evolution in different environments, a careful analysis of the dynamics and the characteristics of galaxies in groups in the local Universe provides the observational data necessary to interpret the different scenarios proposed.

\cite{Tully10} shows that evolved groups, which are characterized by a high fraction 
of 
elliptical galaxies, have at least one  dominant E galaxy and relatively fewer
intermediate luminosity galaxies  than
pristine spiral rich groups \citep[see e.g.][]{Grutzbauch09}. 
The evolution of massive Ellitpicals at the center of groups, expected to be the results of long-term evolution starting from very large scale,
and the evolution of their hot gas content are beginning to be clarified through  self-consistent models
\citep{Bettoni12}.

In this context, we are using a multiwavelength approach to study groups (and their members) characterized by different observational properties (e.g., galaxy density and composition, dynamical properties) to understand whether these are representative of different evolutionary stages or different evolutionary paths.  We have obtained data at several wavelengths through IR, optical and UV observations with e.g., GALEX  and Spitzer \citep{Marino10, Bettoni, annibali11, Marino12, panuzzo}. Here we present the results of XMM-Newton observations of two bright early-type galaxies in poor groups, NGC 4756 and NGC 5328, which we requested in order to 
study the properties of the hot gas in the galaxies and in the groups. Both groups should represent relatively late evolved stages, in which an  elliptical galaxy is the dominant, central member.

The hot gas in the central Elliptical should result both from stellar evolution
(e.g. mass loss from evolved stars) and from
the  environment \citep{Mathews90,Mathews03}, and therefore is expected to be influenced by both the internal and the environmental  properties of the galaxy.
Evolved groups with a dominant  early-type galaxy at their core typically show
hot ($\sim10^7$ K) halos detected in  X rays  \citep[see e.g.][]{mul00,Mulchaey03}, but not all the details of the hot gas properties are clear at the present time. 

We complement the X-ray data with 
a revision of the dynamical properties of the groups, and  with a set of smooth particle hydrodynamics (SPH) simulations which include chemo-photometric models.   These simulations  provide us with predictions for the properties of the system, in particular those of the  stellar population at different stages of evolution, to be compared with the observed  spectral energy distribution 
(SED), and of the hot gas (i.e. luminosity, mass..).

\begin{figure}
\centering
\resizebox{\hsize}{!}{
\includegraphics[clip=true]{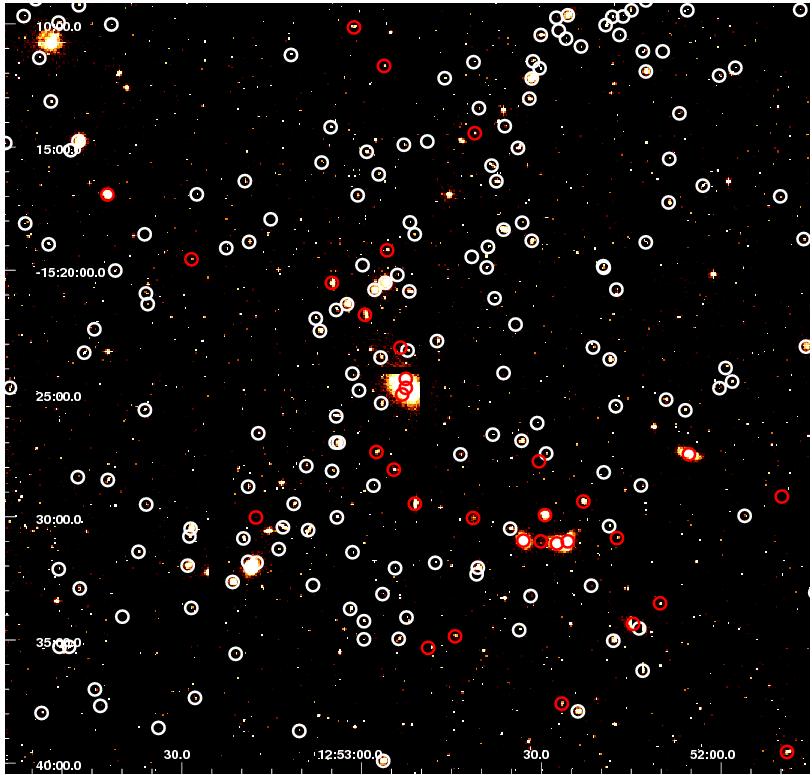}}
\caption{Positions of the galaxies with measured velocities in the  NGC 4756 field, on the un-dithered V-band image of Abell 1631 obtained using 
WFI@ESO-MPI2.2m. Identified galaxies are at the velocity of NGC~4756 (red 
$\sim 4000$ km s$^{-1}$) or at the velocity of A1631 (white, at $\sim 13000$ km s$^{-1}$). Background galaxies are not labeled. 
}
\label{optical}
\label{vels}
\end{figure}

 %
   \begin{figure*}

\includegraphics[width=\hsize]{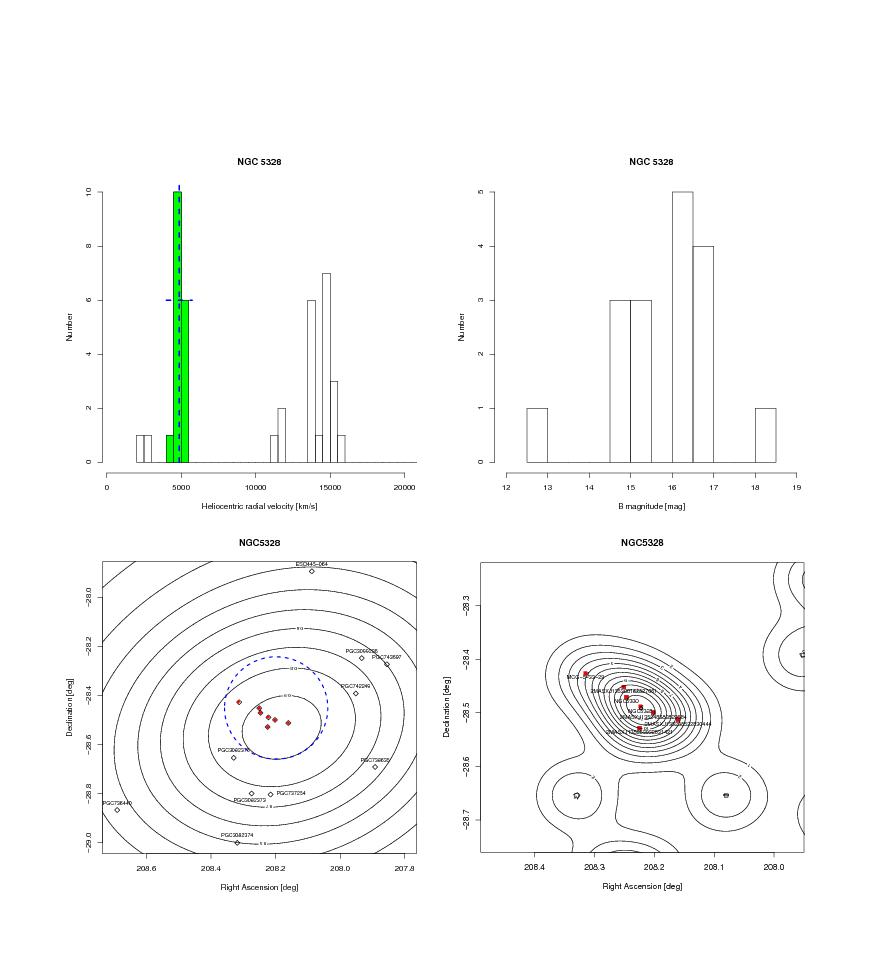}
      \caption{Same as  Fig.~\ref{1pdf} for  NGC ~5328.
Filled red squares are group members listed in \citet{GrutzbaN5328} .}
         \label{2pdf}
   \end{figure*}
%

\begin{figure*}
\centering
\resizebox{\hsize}{!}{\includegraphics[clip=true]{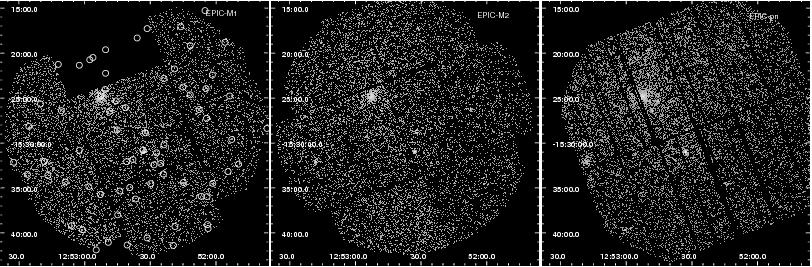}}
\caption{XMM-Newton images of NGC 4756, separately for the three instrument. The positions of the detected sources are shown on the EPIC-M1 image.
}
\label{n4756-fig1}
\end{figure*}

\section{Characterization of  the environment of NGC 4756 and  NGC 5328}
\label{dynanal}

NGC 4756 is the brightest unperturbed elliptical galaxy in a loose
group. The structure of the group is filamentary and  complex, extending
for about half a degree. The central part of the group contains a
significant fraction of early-type galaxies. At about 7' SW of NGC 4756,
a compact, Hickson type, clump of galaxies with signatures of recent
interaction has been identified \citep{GrutzbaN4756}.
The NGC~5328 group, dominated by the homonymous, 
old (12.4$\pm$3.7 Gyr, \citealt{Annibali07}) elliptical galaxy,
has the characteristics of an evolved group \citep{GrutzbaN5328} and is
at the same distance as the Abell~3574 cluster of galaxies, as we explain below.

Both NGC 4756 and NGC 5328 and the groups associated with them  have been extensively studied before by \citet{GrutzbaN4756} and \cite{GrutzbaN5328}. 
However, prior to these observations, inadequate X-ray data were available for both galaxies.  An unpublished snapshot Chandra observation shows emission at the position of NGC 5328, which we will use later to determine the presence of a point source at the center of this galaxy (see \S~\ref{n5328}).  
NGC 4756 was observed by Einstein, the ROSAT HRI and by ASCA. A strong emission peaked on the galaxy was detected by all.  An extension to the N and emission from MCG~--2--33--38, a Sy in the SW sub--group, were also observed.
In spite of the limited quality of the available data, the comparison of  adaptively smoothed images from ASCA in two energy bands  showed an intriguing feature that prompted us to ask for better quality X-ray data. 
The soft ($1-2 $ keV) and hard ($2-8 $ keV)
energy band images from ASCA suggest the presence of two separate
components, with different spatial distributions: a soft peak, consistent with the  hot intergalactic medium associated with
NGC~4756, and  hard band centroid  to the north,
roughly at the position of a concentration of galaxies in Abell~1631  \citep[see][]{GrutzbaN4756}.

Using the much larger spectral and photometric datasets now available in the 
literature,
we first better characterize their environmental properties as follows.

The group around NGC~4756 is well defined in redshift space (see Fig.~\ref{1pdf}, top-left).  
NED lists 38 galaxies at a recessional velocity between 3500-4900 km s$^{-1}$. Of these, 29 are also listed in the {\tt HYPERLEDA} database, which provides us with a  
uniform value of the B magnitude for 27 galaxies, needed to derive the  dynamical properties of the group. We therefore will consider only this subset. One of the objects is PGC043424 (MCG-3-33-17), a bright spiral separated by $\approx$0.7 Mpc from NGC~4756 but with similar recession velocity. \cite{gar93} had included it in the group, but its membership was revised by 
\cite{giu00} and \cite{mah01}, who attribute it to a different group. We therefore consider 26 member galaxies, for which 
we calculate a velocity dispersion of 460 km~s$^{-1}$ at a mean recession  
velocity of 4200 km s$^{-1}$ for the group.

Unfortunately, the group is seen projected onto the background cluster A1361 (the structure seen at $\sim 13000$ km s$^{-1}$ in Fig.~\ref{1pdf}, see Fig.~\ref{vels}) so that a spatial separation of the two structures on the plane of the sky is non-trivial.  This is going to be a challenge in the analysis and interpretation of the X-ray data (\S~\ref{4756X}).

\begin{figure*}
\resizebox{\hsize}{!}
{\includegraphics[clip=true]{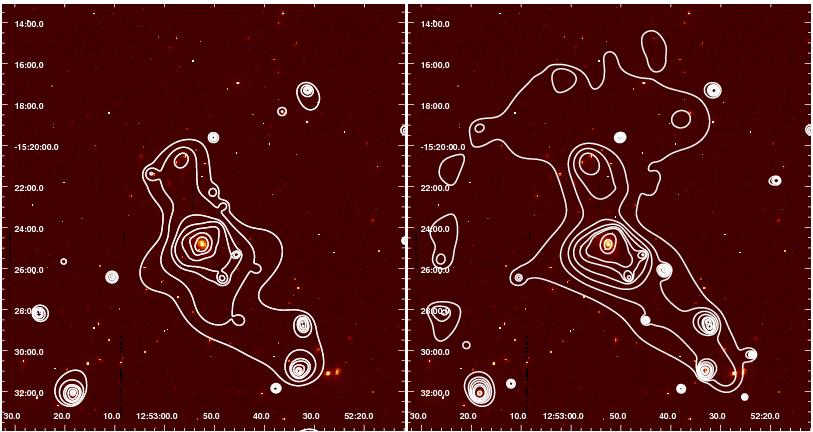}}
\caption{Isointensity contours from the EPIC-MOS2 image treated with an adaptive smoothing algorithm ($csmooth$) in different energy bands: 0.3-1.0 keV (Left) and 1.0-3.0 keV (Right), {\bf superposed onto a DSS-II Red plate.}
}
\label{maps}
\end{figure*}

For NGC 5238, we spatially identify a structure composed of 17 galaxies within a radius of $\sim$700 kpc, with a velocity dispersion of 174 km~s$^{-1}$ at a mean velocity of 4755 km s$^{-1}$ (Fig.~\ref{2pdf} and \S~\ref{dynanal}).  This is located at a projected distance of about 2.5 Mpc NE of the center of the Abell 3574 cluster 
(see \citealt{GrutzbaN5328}).
Given this distance,  the group cannot be considered a cluster substructure at the present time, but the continuity in 
redshift space between 
the cluster ($cz = 4797$ km s$^{-1}$) and the group suggests that 
the two could be potentially connected. 
The  group of NGC~5328 appears thus particularly suited to investigate
processes occurring in galaxy groups which are likely to be accreted by
galaxy clusters.

\begin{table*}
\caption{Dynamical analysis of the NGC~4756 and NGC 5328 groups}
\begin{tabular}{l l c c l c l c c c c c}
\hline\hline
Object  &N. &   \multicolumn{2}{c}{Center  of mass} & V$_{group}$  &  $\sigma$Group      &Dist.$^+$  &  Harmonic   & Virial & Projected & Crossing  & Group \\
   &     &      &  &  & vel.  disp.     &        & radius    &   mass & mass 
       &    time       & $\rm L_B$ \\
    & &   RA.    &        Dec.      & [km~s$^{-1}$]  &   [km~s $^{-1}$]  &  [Mpc]   
 &    [Mpc]      &    \multicolumn{1}{c}{[10$^{13}$ M$_\odot$]}  &   
\multicolumn{1}{c}{[10$^{13}$ M$_\odot$]}  & time$\times$H$_0$    &      [10$^{11}$ L$_\odot$] \\

NGC 4756&26 &193.2462  &-15.4330 & 4155.7 & 445 &  59.37 & 0.10&       2.26 & 17.15 &  0.05 & 0.89\\
NGC 4756$^\dagger$&22&193.2990 &-15.4006 & 4039.61 & 378 &  57.71 & 0.19&    2.97 & 15.49 &  0.05 & 0.63\\
NGC 5328 &17& 208.1982& -28.4508 & 4754.88 
& 174 & 67.93 &  0.10 &   0.34 &  2.23 &  0.09 & 1.24\\
\hline
\end{tabular}

All parameters are weighted  by B-band luminosity using M$_{B\odot}$=5.45 mag\\
$^+$ Distance is derived from V$_{\it hel}$ adopting $H_0$=70 km\,sec$^{-1}$ Mpc$^{-1}$. This gives a scale of 0.302 kpc/$''$ for NGC 4756 and  0.326 kpc/$''$ for NGC 5328.   \\
$^\dagger$ The dynamical anaisys of NGC 4756 is repeated exlcuding the 4 galaxies in the SW clump (see text).
\label{din}
\end{table*}

In Figures \ref{1pdf} and \ref{2pdf} we also show the distributions of the B magnitudes of the member galaxies of both groups. We notice that while the central galaxies NGC 4756 and NGC 5328 are equally bright, the second-brightest members differ by 1.08 and  1.81 mag respectively. 
 Groups with $\Delta$m$>$2 mag between the brightest group member
and the second-brightest member and very luminous in X-ray emission (L$_X> 10^{42}$
erg s$^{-1}$)  are considered evolved ("fossil") groups. The optical properties, in particular  the large $\Delta$m observed between first and second ranked members, would then suggest that the NGC 5328 could be evolved, with properties approaching those of fossil groups. 

The spatial distribution of the member galaxies in the 1.5 Mpc region centered on the brighter member of each group is compared to a 2D binned kernel-smoothed number density 
distribution of galaxies (contours) in the bottom panels of Figures \ref{1pdf} and \ref{2pdf} for both groups. A zoom 
on the inner structure of the group corresponding to the field covered by XMM-{\it Newton} is also shown.
 
The inner, elongated  (P.A. $\approx$ 57$^{\circ}$ NE) density distribution around NGC 4756 presents two cusps,  one in proximity of the galaxy itself, the other generated by a compact group about
7\arcmin\ SW of NGC 4756, originally discussed by \cite{GrutzbaN4756}.  
The inner part of NGC~5328 is composed of a chain of 8
galaxies elongated along P.A. $\approx$ 61$^{\circ}$ NE.

There seems to be a sort of morphology segregation in the group of NGC 5328: among the seven brightest galaxies, two ellipticals, 
including  NGC 5328, occupy the
center of the group (in projection and in redshift space), the
Lenticulars are located from small to medium distances, and the spiral
MCG~--5--33--29 has the largest projected separation from NCG~5328. No
elliptical galaxies are found in the outskirts of the group: the
neighbouring galaxies within 1.5 Mpc are faint spiral or lenticular
galaxies.  

We performed a  dynamical analysis of the two groups following the
luminosity-weighted approach already described in \cite{Bettoni} and
\cite{Marino10}. The dynamical calculations are based on the formulae
given in Table 6 in \cite{Firth06}. The results  are summarized in  Table
\ref{din}. Each galaxy is weighted by its relative B-band luminosity.
The coordinates of the center of mass are obtained averaging the B
luminosity-weighted coordinates of the group members.

The harmonic radii of both groups are the same and their total B
luminosity comparable, but their velocity dispersions and virial masses
are quite different  (see Table 1, lines 1 and 3, and Figures~\ref{1pdf}
and \ref{2pdf}).  We also computed the projected mass for both groups,
which are significantly higher (close to 10$\times$) than the virial mass.
The two estimates should give similar results when the mass is distributed
as the light. In fact \cite{Heisler} note that the virial mass might
underestimate the total mass when the more luminous galaxies are close
in projection and concentrated towards the center.  Projected mass
estimates could be more reliable,  but more sensitive to anisotropies,
subclustering or interlopers.  Discrepancies also arise if groups are in a
transient configuration, i.e.  sub-clumps accreted are not yet virialized.
The relative compact distribution of the galaxies in NGC 5328 might be the
cause of a too small estimate of the total mass from the virial theorem.
This cannot be ascribed to  NGC 4756, for which galaxies are distributed
over a large area. It is possible that this is instead an indication
that the two sub-clumps have not yet reached a stable configuration.
We have then recomputed all quantities for NGC 4756 excluding the 4
galaxies in the SW clump. As shown by the results in Table 1 (line 2),
the harmonic (and virial) radii are significantly different (factor of 2),
the velocity dispersion is reduced by $\sim 25$\% and the masses increased
by a similar amount. Considering the relatively small number of objects
that can be used and the complexity of the field, these estimates should
be used with some caution (see also \citealt{Heisler, aceves, Takizawa}
for similar considerations).

\section{XMM-Newton data}

\begin{figure}
\centering
\resizebox{\hsize}{!}{
\includegraphics[clip=true]{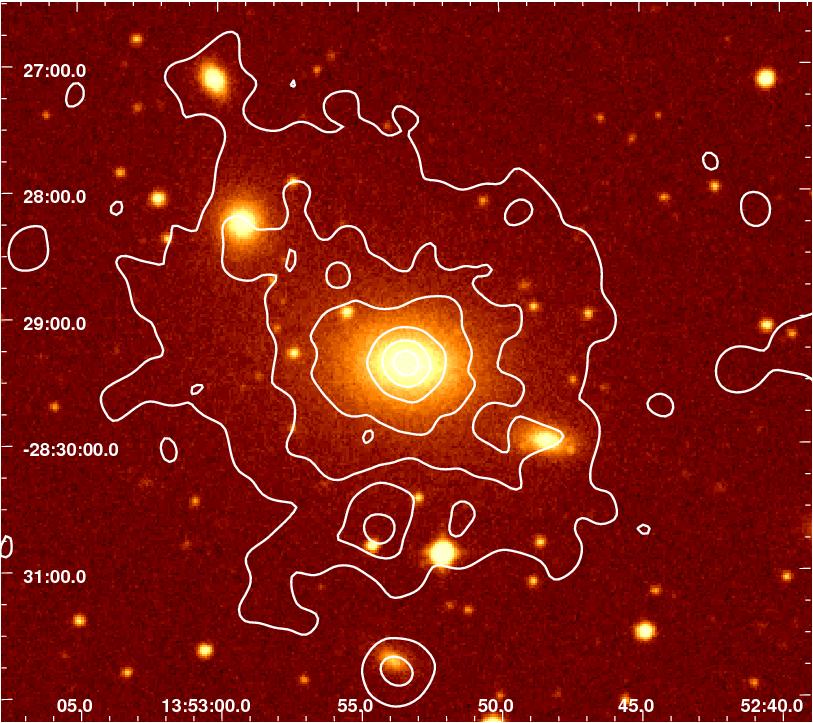}
}
\caption{X-ray iso-contours from the smoothed  0.3-7 keV image of NGC 5328 (from $csmooth$) obtained combining all three instruments
superposed onto the optical 
image from DSS-II Red plate.
}
\label{n5328-fig1}
\end{figure}

\begin{figure*}
\resizebox{\hsize}{!}
{
\includegraphics[clip=true]{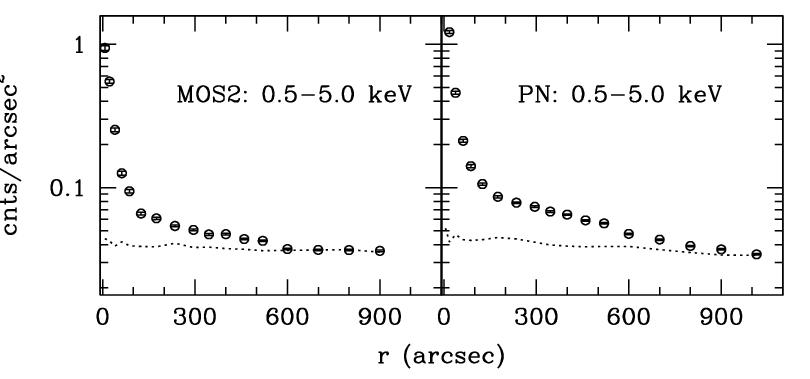}
}
\caption{Raw profiles in
a broad energy bands from clean EPIC-pn and EPIC-M2 data, centered on NGC 4756, azimuthally averaged over 360\degr.
Dashed lines: blank sky appropriately cleaned and sky-casted, produced from observations in  neighbouring fields, and normalized to match the data at large radii (see text)
}
\label{radprof}
\end{figure*}

We obtained a $\sim$28ks XMM-Newton observation of NGC 5328 with EPIC instruments 
between January and February 2007. In December
2008 we obtained a $\sim$70ks XMM-Newton observation of NGC 4756. The EPIC detectors operated in PrimeFullWindow mode with Thin
filter. We analyzed the data with XMM-Newton Science Analysis Software
$xmmsas\_ 20100423\_ 1801$
 using the standard data reduction and the latest calibrations as
suggested in the ``User's Guide to the XMM-Newton Science Analysis
System''\footnote{xmm.esac.esa.int/external/xmm\_user\_support/documentation/
sas\_usg/USG/}.  We cleaned the data from periods of high flares applying
a filter on the count rate derived from the light curve provided with
the standard processing (5 for EPIC-pn and 2 for EPIC-MOS) and obtained a
clean time  of 12.7 ks and 10.4  ks for EPIC-pn  and for each EPIC-MOS respectively for NGC 5328 and $\sim 24/48$ ks  for NGC 4756.
We also 
identified bad pixels, bad columns and CCD gaps which we excluded from
the analysis.  

\subsection{Imaging analysis}

We created binned X-ray images separately for the three
instruments in different energy bands for both galaxies, and a combined image from all instruments for NGC 5328 (see later and Fig.~\ref{n5328-fig1}). 
Since the analysis of NGC 4756 involves non-standard procedures, and given that, unlike the case of NGC 5328, the non-operational status of one of the CCDs in EPIC-M1 is affecting heavily our analysis, 
we do not consider MOS1 data for NGC 4756.
The raw images of the three instruments are shown separately in  Fig.~\ref{n4756-fig1}.

A better representation of the emission in the field is given by
Fig.~\ref{maps}, where we have smoothed the EPIC-M2 data with an adaptive
smoothing algorithm (using the task $csmooth$ in the $CIAO$ software, see
\citealt{fru}) in different energy bands.  The data are not corrected for
exposure or background at this stage, so they provide a first indication
of the location of the emission, to guide the subsequent analysis.  It is
clear from the figure that the peak of the emission is centered on NGC
4756, but there is significant extended emission to the north and SW, and
that there is a secondary peak to the N, more evident in the harder band,
at the position of two bright galaxies belonging to the  cluster A1361.

For NGC 5328 we  applied an adaptive smoothing algorithm with a Gaussian
filter to images obtained combining the three instruments in different
energy bands. The results are shown in Fig.~\ref{n5328-fig1} for the
broad band, as a contour plot superposed on the optical DSS plate. The
region  where the emission is detected is all within the central CCD
and  does not extend to the missing CCD in EPIC-MOS1.

\begin{figure*}
\resizebox{\hsize}{!}
{
\includegraphics[clip=true]{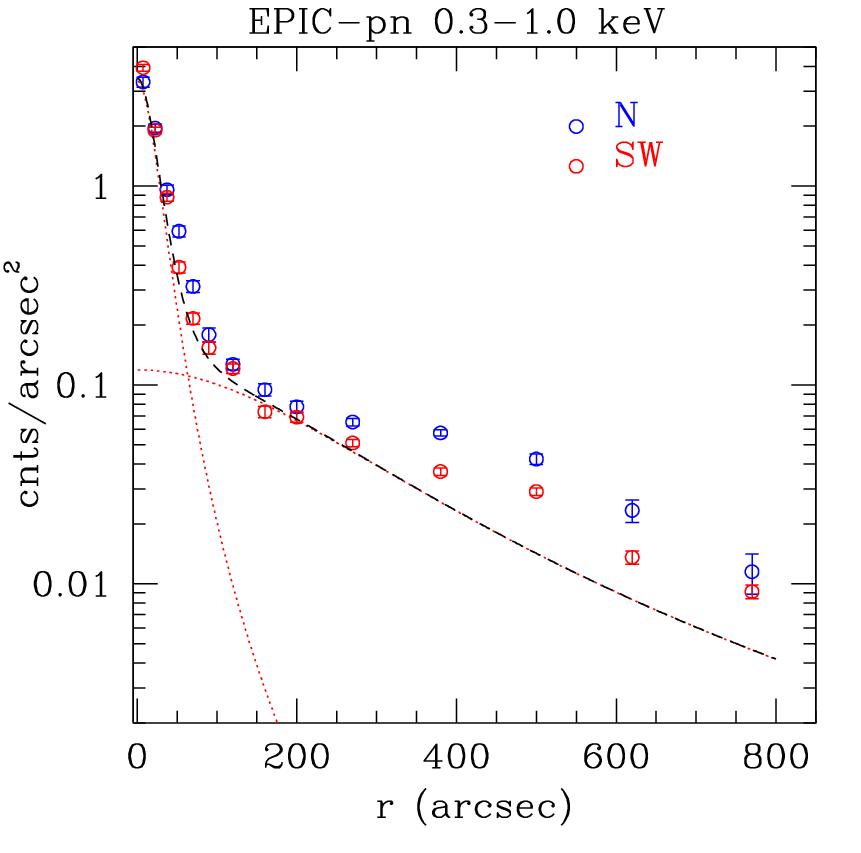}
\includegraphics[clip=true]{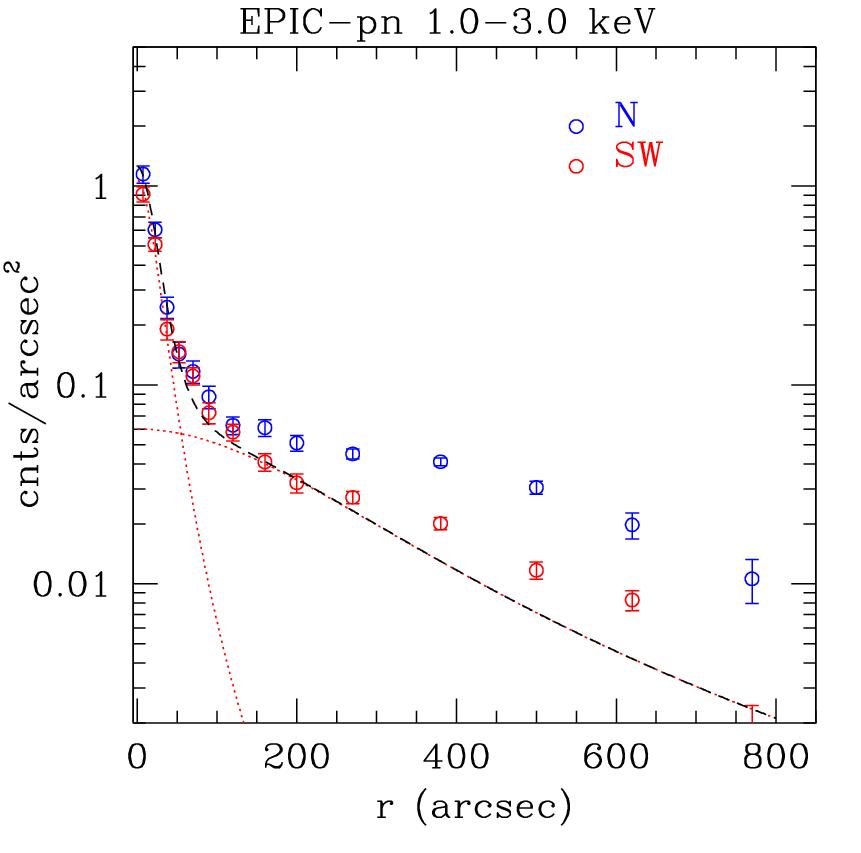}}
\caption{Azimuthally averaged net profiles in the 0.3-1 and  1-3 keV band from clean pn data. Center on NGC 4756, azimuthally averaged 
between $\sim 315-70\degr$ (N) and $\sim 140-315$ (SW). Background is from the normalized blank sky profiles in the same region. The dashed line indicates model composed of   two $\beta$-models with: r$_c$=$33''$;$\beta$=0.9  and  r$_c$=300$''$; $\beta$=0.7
}
\label{net}
\end{figure*}

To identify individual sources in the field of each galaxy, we
ran a detection routine on the
cleaned event files of the three instruments at the same time
in the 0.5-4.5 keV band (see \S~\ref{appendix}).  
Besides the target, we detect a total of 68 sources in the NGC 4756 field, listed in Table~\ref{n4756-tab}   and 32 in the  NGC 5328 field (Table~\ref{n5328-tab}). Possible identification with known objects are given.

We use these lists to identify individual sources of interest related to the target galaxies, and to exclude sources from subsequent analysis [with standard radius of 30$''$]. 
We extracted radial profiles in different energy bands and 
spectra in different regions together with the background, ancillary response (ARF) and detector response matrix (RMF) files  for the three instruments separately, and analyzed the results with XSPEC version 12.

In what follows we describe in more details the analysis of each target separately, since the two objects require  different ad-hoc procedures.

\subsubsection{NGC4756}
\label{4756X}

As shown by Fig.~\ref{n4756-fig1} and Fig.~\ref{maps}, a bright source is visible towards
the northern edge of the central EPIC-MOS CCD, coincident with NGC
4756. A few other sources are also present in the
field of view:  to the SW, there are a few coincident with galaxies in
the southern subclump; an extension to the N coincides with galaxies belonging to the background cluster A1361.

We therefore have used two different approaches for this source: we have examined the data assuming the center of the emission on a) NGC4756 and b) on the secondary peak to the North, with the aim of disentangling the emission from the two components -- foreground group and background cluster.  We report here the analysis step-by-step and will summarize the results in \S~\ref{results}.
We extracted radial profiles from both EPIC-pn and EPIC-MOS centered on the peak of the emission on NGC~4756
in different bands (we show 
the broad band profiles in Fig.~\ref{radprof}), azimuthally averaged over 360\degr,  and in different regions of the field of view.  
As a first approach we extracted profiles separately for the North and South half of the plane in circular concentric annuli of increasing width with radius, and for the 4 NSEW directions. 

All profiles appear to have a radial gradient out to r$\sim$600$''$, with the exception of the E direction that traces a smaller range, indicative of emission above the expected background at least out to these radii. However, we need to take into account the fact that the profiles trace very different regions of the detector, due to the off-center position of the source.  We therefore requested blank sky images  appropriate for this field from the XMM-Newton sky background
team\footnote{http://xmm.vilspa.esa.es/external/xmm\_sw\_cal/background}, which we use to estimate the shape of the profile of the background emission in the same regions.

The selection criteria for the blank sky fields we found most appropriate for the success of our analysis were based on revolution, column density and spatial direction, with the additional constraint on the presence [rather absence] of detected sources in the fields.  More specifically only fields taken with the following constraints were considered: a) only revolutions subsequent to the loss of CCD6 in EPIC-MOS1 (revolution 961 and after); b) galactic column densities N$_H$ in the range $ 1.5-4.5 \times 10^{20}$ cm$^{-2}$; c) pointings within a radius of 90\degr\ from the direction of NGC4756; d) no more than 2 sources in the region between NGC 4756 and the southern sub-condensation,  and none with a flux f$_X > 6 \times 10^{-13}$ erg cm$^{-2}$ s$^{-1}$ (0.2-12.0 keV). The reference list used was the second XMM-Newton source catalog. The final blank fields  have been cleaned using the same criteria as the original files.

To normalize the blank sky fields to the data we have first rescaled the background for the relative exposures, and then applied a small correction to better match the light distributions.  The correction is of the order of 10\%. The resulting radial profiles are also shown in Fig.~\ref{radprof}. 

Comparison between the profiles in the N/S halves in different bands 
suggests a systematic excess best seen in the intermediate 1-3 keV energy range in the N half at $\rm r> 150''$. Little difference is seen between the profiles obtained in the S and W directions.  We therefore refined our choices of regions, so that the N sector is included between $\sim -45\degr$ and $\sim 70\degr$ (counterclockwise, from N), and the SW is between $\sim 140-315\degr$. 

We have then produced background-subtracted profiles shown in Fig.~\ref{net} for the 0.3-1 and  1-3 keV band, obtained using the rescaled blank-sky profiles from the same regions in the same energy bins.
It is evident that the level of emission is different in the two directions, higher towards the north outside  r $\sim 150''$, and most evident in the 1-3 keV range. The  emission extends in both directions out to the edges of the field of view.

A second difference between the profiles in the two bands is the relative importance of the central peak to the more extended emission.  This can be seen from the comparison of the $\beta$-type models we plot on the data.  Note that these are not fits to the data, although they could be representative of the emission out to radii of $250''-300''$ in the SW direction. While the overall normalization is smaller in the 1-3 keV band, relative to the 0.3-1 keV, we had to reduce also the relative normalization between the inner and outer models, by a factor of $\sim 1.5$. This could be indicative of two distinct components, one dominating below r$\sim 50''-70''$, with different spectral characteristics.  This will be further discussed in section~\ref{sec:spec-analysis}

\begin{figure}
\resizebox{\hsize}{!}
{
\includegraphics[clip=true]{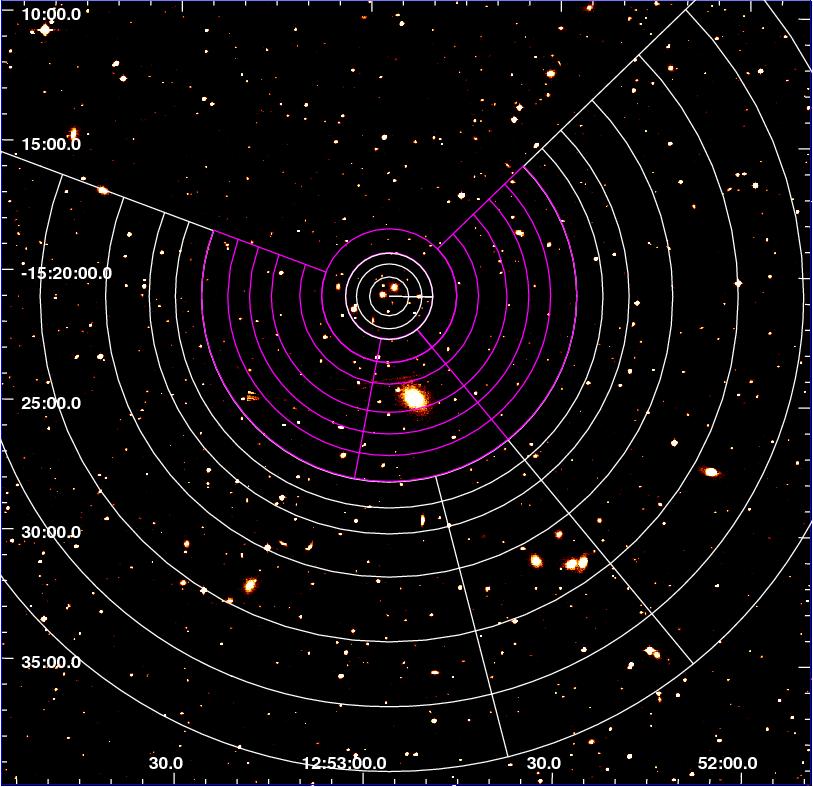}}
\caption{Regions used to derive the radial profiles shown in Fig.~\ref{netoncl}. The annuli in the innermost region are azimuthally averaged over 360\degr, in the outer regions the angles are chosen both to accommodate the limits of the field of view and to maximize the emission in the S direction towards MCG-02-33-038.
}
\label{directions}
\end{figure}

\begin{figure*}
\resizebox{\hsize}{!}
{\includegraphics[clip=true]{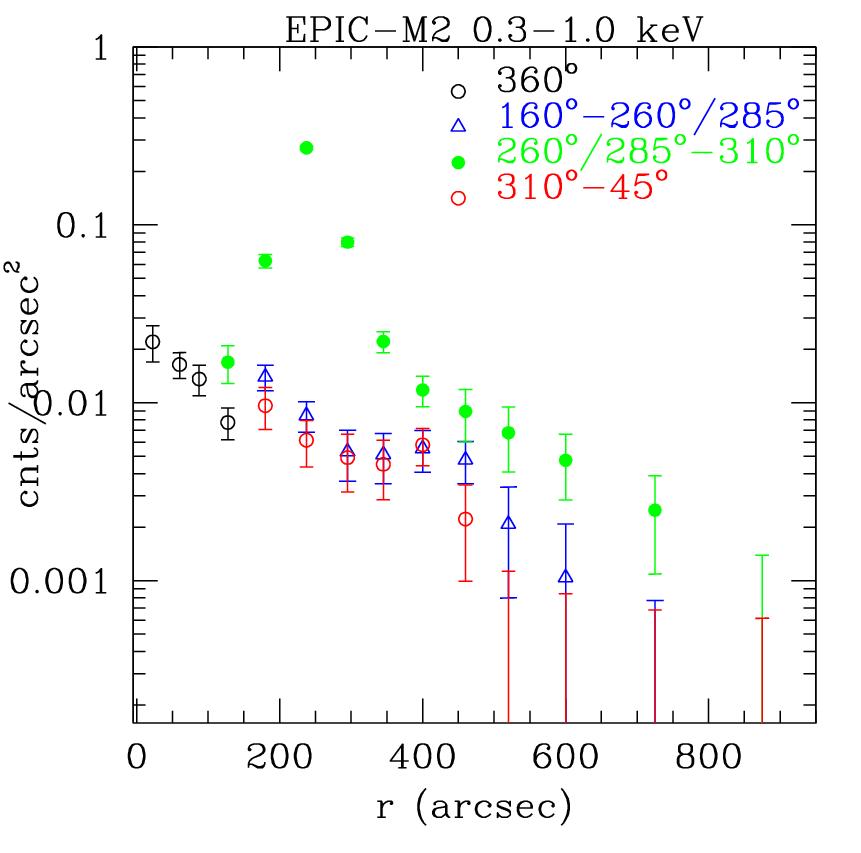}
\includegraphics[clip=true]{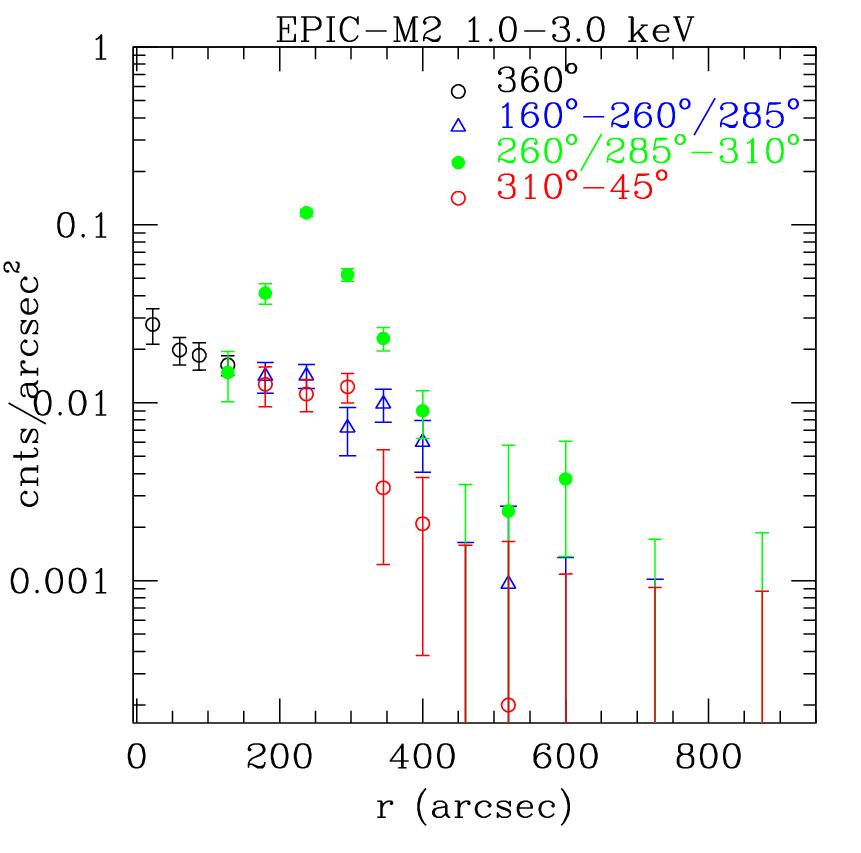}}
\caption{Azimuthally averaged net profiles in the 1-3 keV band from clean pn data. Center on NGC 4756, azimuthally averaged over 180\degr.  Background is from the normalized blank sky profiles in the same region. The northern half shows excess starting at $\sim 100''$ radius.  
}
\label{netoncl}
\end{figure*}

Given the evidence of emission in all directions, and stronger to the N, we have moved the center of the profiles onto A1361, to better define the contribution from the cluster.  This is even further out than NGC~4756, outside the central  CCD in the EPIC-MOS, close to the edge of the field.  
We have considered different directions to derive radial profiles, and we resolved on the configuration shown in Fig.~\ref{directions}.  Given the off-axis location of the cluster, we could use 360\degr\ annuli only in the innermost region, before getting too close to the edge of the field. We have then chosen a preferred direction to maximize the contrast and better highlight the emission from the galaxy and group relative to that of the cluster.  In Fig.~\ref{netoncl} we present the resulting net profiles above the background that we have estimated from the blank sky fields used above.  We show the EPIC-M2 data, since it has fewer  CCD gaps, and none between NGC4756 and the southern subclump of interest (see later).  We could not use EPIC-M1 since the cluster would fall on the lost CCD.

The profiles we obtained indicate that: 1) there is a general decrease of the emission centered on A1361, consistent with emission from a cluster, out to a radius r$\sim 400''- 600''$ in all directions we could probe; 2) the contribution from NGC4756 is evident at radii greater than  150$''$, in the profile obtained in the S-SW direction; 3) the central regions of the 1-3 keV profile indicate stronger emission than in the softer band, unlike the profiles obtained for NGC 4756, indicating a rather different spectrum; 4) the softer profile shows an excess out to $\sim 800''$ in the S-SW direction, relative to the profiles obtained in the contiguous regions, while a possible excess in the harder  band is less convincing.  This confirms that there is  emission from hot gas on NGC 4756 in excess of what is expected from the background cluster,  and indicates that the emission extends in the direction towards MCG-02-33-038.

\begin{table*}
\caption{Detected members. 
\label{members}}
\begin{tabular}{llrrrrll}
\hline 
\hline
Object & morphology &log L$\rm _b$&log  L$\rm _k$&Radius&\multicolumn{1}{c}{Net counts}& \multicolumn{1}{c}{ $f_x$ } &\multicolumn{1}{c}{ L$_x$}\\
& &&&[X rays]&0.5-4.5 keV& 0.5-2~;~2-10 keV & 0.5-10 keV \\
& &&&kpc \ \ &\multicolumn{1}{c}{EPIC-pn}&\multicolumn{1}{c}{\ergcms} &\multicolumn{1}{c}{\ergs} \\
\hline
\hline
&&&\\
\multicolumn{8}{c}{NGC 4756 group} \\
\hline \hline
NGC 4756$^\dagger$ &E3 &10.43 & 11.23& 40 &6038.2$\pm$119.8 &4.8~~;~~0.2 $\times 10^{-13}$& 2.1$\times 10^{41}$\\
\hline
IC0829 &SB0&10.02&10.82&4.5& 67.9$\pm$11.0&4.7~~;~~8.5 $\times 10^{-15}$&5.7$\times 10^{39}$\\
MCG-02-33-038$^{\dagger}$ &S0/a&9.81&10.50&9 &871.0$\pm$ 33.2 &7.4~~;~~7.2 $\times 10^{-14}$ & 6.3$\times 10^{40}$ \\
\hline
\hline
&&&\\
\multicolumn{8}{c}{NGC 5328 group} \\
\hline
\hline
NGC 5328$^\dagger$ &E1 & 10.89&11.58 & 110 &1906.5$\pm$54.8 & 6.4~~;~~0.6 $\times 10^{-13}$& 3.9$\times 10^{41}$\\
\hline
PGC3094716&E&9.57&10.35 &4.9&85.3$\pm$20.5& 4.9~~;~~8.9 $\times 10^{-15}$ & 7.7$\times 10^{39}$  \\
2MASX J13525393-2831421& S0/SB0& 9.12&9.82& 4.9&77.4$\pm$20.3 &4.4~~;~~8.1 $\times 10^{-15}$&   7.0$\times 10^{39}$\\
NGC 5330&E/E1&10.07&10.74 &4.9&292.4$\pm$25.0 &1.7~~;~~3.1 $\times 10^{-14}$&   2.6$\times 10^{40}$ \\
ESO 445- G 070 &SB(r)0/a&9.84&10.63 &4.9&\ \ 30.0$\pm$9.0$^{\dagger\dagger}$& 3.3~~;~~6.1 $\times 10^{-15}$&   5.2$\times 10^{39}$ \\
\hline
\end{tabular}
\\
\\

Data from NED and EPIC-pn. K-band luminosities assume M$_{K\odot}$=3.28.  X-ray fluxes and luminosities are computed assuming a power law model with $\Gamma=1.7$ and
line-of-sight absorption of 3.5 and 4.6 $\times 10^{20}$ cm$^{-2}$, for NGC 4756 and NGC 5328 respectively  \citep{LAB}.\\
$^\dagger$ X-ray quantities derive from the spectral analysis, assuming a plasma model for NGC 4756 and NGC 5328, and power law  with $\Gamma=2.1$ for MCG-02-33-038.  For NGC 4756 flux and luminosities refer to the soft (kT$\sim$0.7 keV) component only. \\
$^{\dagger\dagger}$ On a gap in EPIC-pn; counts are derived from the sum of the 2 EPIC-MOS
\end{table*}

Two other galaxies in the NGC 4756 group are detected: IC0829 and  MCG-02-33-038 in the southern sub-clump (see Appendix and Table~\ref{n4756-tab}). Table~\ref{members} summarizes the characteristics of these galaxies, together with their fluxes in a soft and a hard energy band. We have re-evaluated their X-ray luminosity above a local estimate of the background, which takes into account possible local diffuse emission.   We have extracted the source counts from EPIC-pn in  a circle of $20''$ and $30''$ respectively.  For 
MCG-02-33-038 we have enough counts to fit the spectrum with a more appropriate model (a slightly absorbed power law with $\Gamma=2.1$, see next section), that reflects the presence of a low luminosity nuclear source. IC0829 does not have enough counts for a spectral analysis, so we have used the standard model, $\Gamma=1.7$ and Galactic absorption, to evaluate the flux. The luminosity of IC0829 is entirely consistent with what could be expected by the LMXB population in a galaxy with the same K-band luminosity \citep{boroson}.

\subsubsection{NGC 5328}
\label{n5328}

Figure~\ref{n5328-fig1} shows  extended emission clearly peaked on NGC 5328. A slight  elongation towards NGC 5330 to the NE is visible.  NGC 5330 itself is also detected, though marginally, as confirmed by the source detection algorithm, which detects 3 additional members.  Table~\ref{members} gives some of their basic properties, together with  the net counts (with errors) from EPIC-pn, the
flux and the luminosity. Fluxes are obtained assuming a simple
spectral representation with a power law with fixed slope at $\Gamma$= 1.7   and Galactic N$_H$.  Luminosities are of the order of $10^{40}$ \ergs, typical of normal early type galaxies. 

\begin{figure*}
\resizebox{\hsize}{!}{\includegraphics[clip=true]{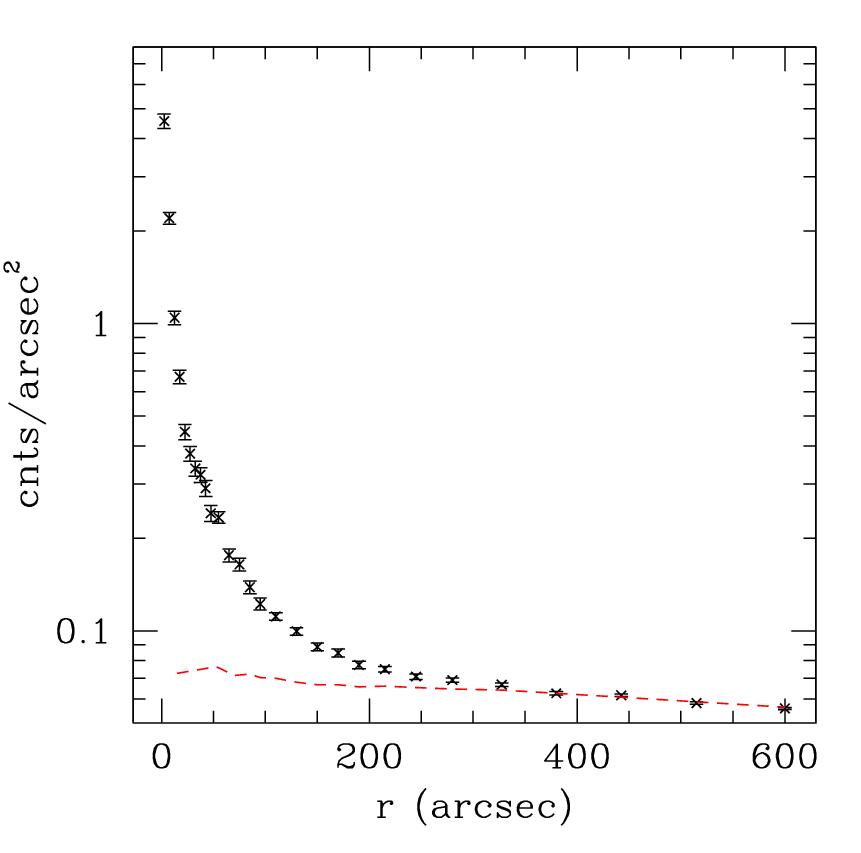}
\includegraphics[clip=true]{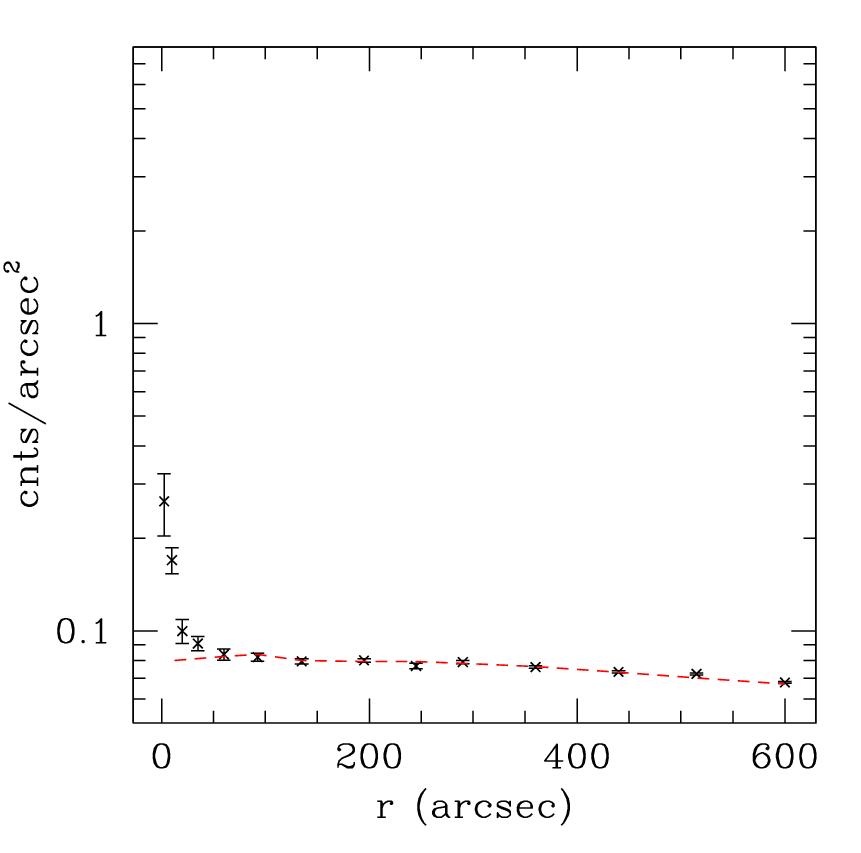}
}
\caption{Azimuthally averaged raw radial profile of the emission from NGC 5328 observed with EPIC-PN in 0.3-2 keV (left) and 2-7 keV (right) bands. The red line is the shape of the blank sky profile arbitrarily normalized to the data. }
\label{n5328-fig2}
\end{figure*}

We extracted radial profiles from the emission of NGC 5328 using concentric annuli centered on the X-ray 
peak. We used the mask files and source regions produced before to exclude ``dead columns'', gaps and 
detected sources. The
radial profiles extracted from the EPIC-pn 
in the 0.3-2 keV and 2-7 keV bands are shown in Figure~\ref{n5328-fig2}. 

It is immediately evident from the figure that the emission is extended in the soft band, while at harder energies the emission is limited to the innermost region, at r $<50''$, and then becomes almost constant with radius.

To better evaluate the total extent of the source, we compared the
radial profile obtained from the NGC5328 with that obtained from the XMM-Newton blank sky fields (see above).
We extracted this latter from event files cleaned using the same
procedure used for NGC 5328 data, rotated and sky-casted to match
the original data, from the same region file centered at the same
position.  We notice that the shape of the blank sky profile,
arbitrarily renormalized to match the data at large radii
provides in
fact a good representation of the profile for radii r$>300''-350''$. 
We therefore use this radius as the maximum extent for this
source. This corresponds to a physical radius of $\sim$100 kpc.
The comparison in the hard band confirms that the emission
extends only out to r$\sim 50''$.

The net profile of the emission in the 0.3-2.0 keV band is
shown in Fig.~\ref{n5328-net}.  In the same figure we also plot a combinations of two $\beta$-type models that could represent the radial distribution of the emission (as for NGC 4756, these are not fits).  In this case a single model could represent the data with the exception of a small excess in the center, at r $< 30''$.

To better understand the origin of this excess we checked the Chandra image for the existence of a central point source. The very short exposure time collected about 250 counts (0.5-2.0 keV band) in a region of 40$''$ radius centered  on the galaxy.   The radial  distribution of the photons indicates that the source is extended, at least as the optical extent of the galaxy.  A crude estimate  of the flux based on a APEC MODEL indicates  a flux consistent with XMM-Newton in the same region.   It is therefore likely that the excess in the XMM-Newton data is due to the emission from the galaxy itself, while a group component contributes to the emission at larger radii.

\subsection{Spectral Analysis}
\label{sec:spec-analysis}

We used a local background, extracted from the same datasets (rather than from the blank sky fields) in well chosen regions appropriate for our purposes. This ensures that we are indeed obtaining the most representative background for each region we study, although in some cases small corrections for different responses in different parts of the detector might be appropriate.  Considering that the source regions are relatively compact and centrally located in NGC 5328, and given  the complexity of the field around NGC~4756, we feel that these are significantly less important than any correction that would be required from using blank sky fields for background.
We present the results we obtained using the EPIC-pn data, but they were checked against the EPIC-MOS data in equivalent regions.  We have used a combination of APEC and power law models with line-of-sight absorption at low energies \citep[3.5 and 4.6 $\times 10^{20}$ cm$^{-2}$,][ for NGC 4756 and NGC 5328 respectively]{LAB}.  For the APEC, we considered metal abundances of 50-100\% the cosmic value, using the  abundance tables from \cite{angr}.
To improve the statistics, we have binned the spectral data: depending on the statistics, we have  binned to include  30-60 counts per bin or to reach a 2$\sigma$ level  after background subtraction.

\subsubsection{NGC4756} 

The source associated with NGC4756 is clearly projected onto a more diffuse emission from the background cluster.  We have therefore identified a few regions where we can safely isolate NGC4756 as the dominant component, and used the neighbouring regions, that include the cluster emission, as background.

The X-ray peak on NGC 4756 and surrounding regions are clearly well above the expected cluster emission, so we have selected a region of 45$''$ radius (corresponding to the major axis size of the galaxy) and the adjacent annulus with outer radius of 135$''$ (close to the maximum extent seen from Fig.~\ref{netoncl}) to characterize the emission from NGC4756 and immediate surrounding. 

We have also selected a circle of 75$''$ radius, with the exclusion of the Seyfert nucleus, to characterize the southern compact group, and an ellipse along the  connection between NGC 4756 and the southern group (see Fig.~\ref{maps} and Fig.~\ref{netoncl}) to trace the emission between these two systems. 

We find that the spectrum of NGC 4756 is well described by an APEC model with kT$\sim 0.64 \pm 0.02$ keV and N$_H\sim 2 \times 10^{20}$ in excess of the galactic value (but consistent within the 90\% errors with the line of sight value), and luminosity of  $\sim 1.3 \times 10^{41}$ erg s$^{-1}$ (0.5-2.0 keV).  The data do not require a power law component, but can accomodate one  with index fixed at the value  $\Gamma$=1.7, which contributes a luminosity of  $\sim 3.3 \times 10^{39}$ erg s$^{-1}$ (0.5-2.0 keV) or  $\sim 9.3\times 10^{39}$ erg s$^{-1}$ in the 0.3-8.0 keV band.  
This is actually slightly lower (but consistent within errors) than the contribution from LMXB sources estimated by \cite{boroson} based on the K-band luminosity. 

\begin{figure}
\resizebox{\hsize}{!}{\includegraphics[clip=true]{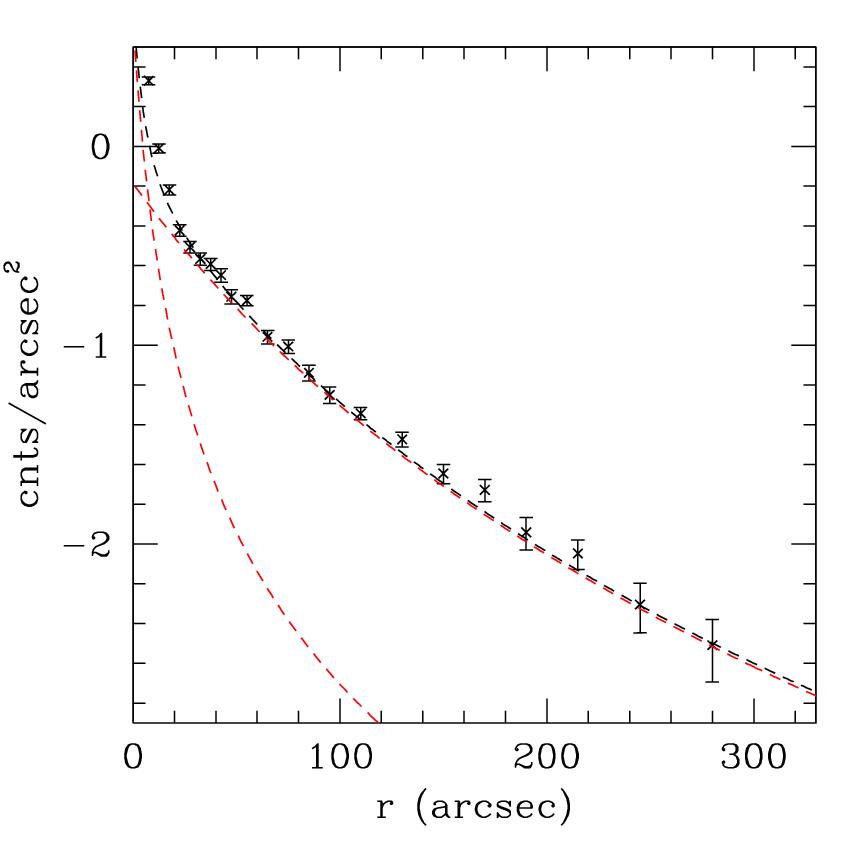}}
\caption{Azimuthally averaged net profile of the emission from NGC 5328 observed with EPIC-PN in 0.3-2 keV. The dashed line represent a model given by the sum of two $\beta$-models with: r$_c$=$6.6''$;$\beta$=0.63 (PSF) and  r$_c$=160$''$; $\beta$=1.05}
\label{n5328-net}
\end{figure}

The surrounding region has kT$\sim 0.76\pm0.03$ keV, with luminosity of  $\sim 5 \times 10^{40}$ erg s$^{-1}$ (0.5-2.0 keV).  In this dataset, there is a requirement of a power law component with L$_x \sim 2.1 \times 10^{40}$ erg s$^{-1}$ (0.5-8.0 keV). An equivalently good fit is obtained by using a second APEC component, with kT$\sim 3.6$ keV (kT$>1.5$ keV) and 50\% abundances.  This could represent a second ``group component'' or some contribution from the background cluster (at  L$_x \sim 1.4 \times 10^{41}$ erg s$^{-1}$, 0.5-2.0 keV, at the distance of the cluster).  

Similar values are observed in the extended emission in the Southern compact group. The mean temperature is kT$\sim 0.68$ and a luminosity of L$_x \sim 3.6 \times 10^{39}$ erg s$^{-1}$. 
We also found a slightly better minimum at  kT=0.2 keV, but this would also require a significantly higher absorption N$_H \sim 6 \times 10^{21}$, which is not observed in the neutral hydrogen.
A power law with (fixed) $\Gamma=1.7$ is needed to explain the high energy tail, with a contribution of 
L$_x \sim 6.3 [10.4]\times 10^{39}$ \ergs\ [0.5-2;2-10], to account for excess at high energies.     
This is unlikely due to the contribution of the Sey nucleus, which is not included in the spectral region, but which has a steeper spectrum with $\Gamma=2.1 \pm0.2$ 
and moderate intrinsic absorption (N$_H \sim 4\times 10^{20}$, $<8 \times 10^{20}$ cm$^{-2}$ at the 90\% confidence; see also Table~\ref{members}). 

\subsubsection{NGC 5328} 

We extracted the X-ray spectrum of the target in a circle with 
radius r=40\arcsec{}, centered on NGC 5328,  and in the adjacent annulus of radii  r=40\arcsec{}-350\arcsec{}. 
The background is derived from a set of source free circles outside the extended emission.

We used an APEC model to represent the hot gas.
We fix the Galactic absorption to the line of sight value
of $N_H\sim$4.6 $\times 10^{20}$ cm$^{-2}$.

The temperatures in the inner and outer regions are very similar, although 
formally different:
kT$=0.85\pm0.2$  and kT$=0.95\pm0.4$ (inner/outer respectively), indicative of a slightly smaller temperature towards the center.
The $\chi^{2}_{\nu}$ is in the range of  1.1-1.4  (for 304 DoF).
There is no need for a power law contribution, expected from binaries, which however is expected to be well below the plasma component.
The gas  has intrinsic
luminosities of L$_X = 9.81 \times 10^{40}$ \ergs\ and L$_X = 2.56\times 10^{41}$ \ergs\ (0.5-2 keV) in the inner and outer regions respectively.

\section{Results from the X-ray data}
\label{results}

The spatial analysis of the fields of NGC 4756 and NGC 5238 indicates that both galaxies sit at the center of a large halo of hot gas.  The shape of the emission around NGC 5328 suggests a rather relaxed, small group with total size of $\sim110$ kpc radius, and luminosity L$_x \sim 3.5 \times 10^{41}$ \ergs.  

The presence of the background cluster onto which NGC 4756 is projected prevents us from determining the shape and the true extent of the emission, although we have evidence that there is a connection with the sub-group to the SW, at $\sim 130$  kpc. The luminosity that we can attribute to this system is L$_x \sim 2 \times 10^{41}$ \ergs\ within $\sim 40$ kpc, and L$_x \sim  10^{40}$ \ergs\ from the southern clump and the intermediate region.  This  is probably an underestimate of the total emission from this system as a whole, since we have only measured the emission from selected regions, where the contamination from the background cluster can be quantified.

In both cases a soft thermal plasma with a temperature around 0.6-0.9 keV is detected at the center, with somewhat higher values towards the periphery, consistent with what is observed in other groups of galaxies \citep[e.g. NGC 4261][and references therein]{osull}.

A few member galaxies are detected, besides the central ones, with luminosities of about L$_x \sim  10^{40}$ \ergs, in line with the X-ray luminosities of the LMXB component in normal, early type galaxies of that optical/IR luminosity.  A higher luminosity L$_x  \sim 6 \times 10^{40}$ \ergs\ is observed in MCG-02-33-038, to the SW of NGC 4756, due to a slightly absorbed Sey2 nucleus. The other Sey2 nucleus, ESO445-G070 in the NGC 5328 group, has a lower luminosity than  MCG-02-33-038, but higher than the other objects and marginally higher than expected from the LMXB contribution in early-type galaxies \citep{boroson}. NGC 5328 is $\sim1'$ from the peak of the emission and NGC 5328, and is detected with too few counts for a proper spectral investigation, but we cannot exclude that the higher value is due to a small contribution from the nucleus itself, or from a small gaseous component.

Within the  limitations of current statistics, we lack evidence for a gaseous  component in the detected members of either groups.  This is in fact different from what \cite{Jeltema08} found in their analysis of 13 nearby groups, where  they detect emission from gas in a large fraction of the objects, although with lower luminosity than in field systems of the same K-band luminosity.  \cite{Jeltema08} interpret this as a result of mild ram pressure, which acts on the galaxies although less efficiently than in clusters. Taken at face value, this result would suggest that stripping has been even more efficient in these two groups than in the 13 studied by Jeltema.  The luminosities of the extended emissions in  NGC 4756 and NGC 5328 are comparable to relatively faint systems in Jeltema, such as HCG 42 and NGC 3557,  which also have comparable or higher velocity dispersions ($\sim 300$ km s$^{-1}$), so we could expect a similar action from the ambient medium.  Three objects have been detected in NGC 3557 at luminosities L$\rm _{x,gas} \sim 4-40 \times 10^{39}$ \ergs\ (0.5-2.0 keV), while no detections are found in HCG42, with limits at L$\rm _{x,gas}\simeq 1 \times 10^{39}$ \ergs.  The higher spatial resolution of the Chandra data used by Jeltema could contribute to their higher success in measuring hot gas.  However, the K-band luminosities of the galaxies in NGC 4756 and NGC 5328 are lower than those of the detected galaxies in Jeltema (see their figure 1), so we cannot exclude that our failure to detected hot gas in them is due to the significantly lower values expected.

We notice that the X-ray luminosities we detect are relatively low for groups, but consistent with the tail end of the distribution.  If we attribute it all to the central galaxy, the size of the detected emission is significantly larger than the stellar distribution, so it is reasonable to expect a contribution from a larger potential. 
If we consider the inner regions studied in the spectral analysis, or the inner $\beta$-model component (see Figs.~\ref{net} and \ref{n5328-net}),  as due to the galaxy proper, then the emission is co-spatial with the stellar light ($\leq D_{25}$). 
This is also consistent with the [marginal] evidence  that the inner region is softer, and could be interpreted as due to a sizable contribution from the gas produced by stellar sources and retained within the optical extent of the galaxy. 

\begin{figure*}
\includegraphics[width=\hsize]{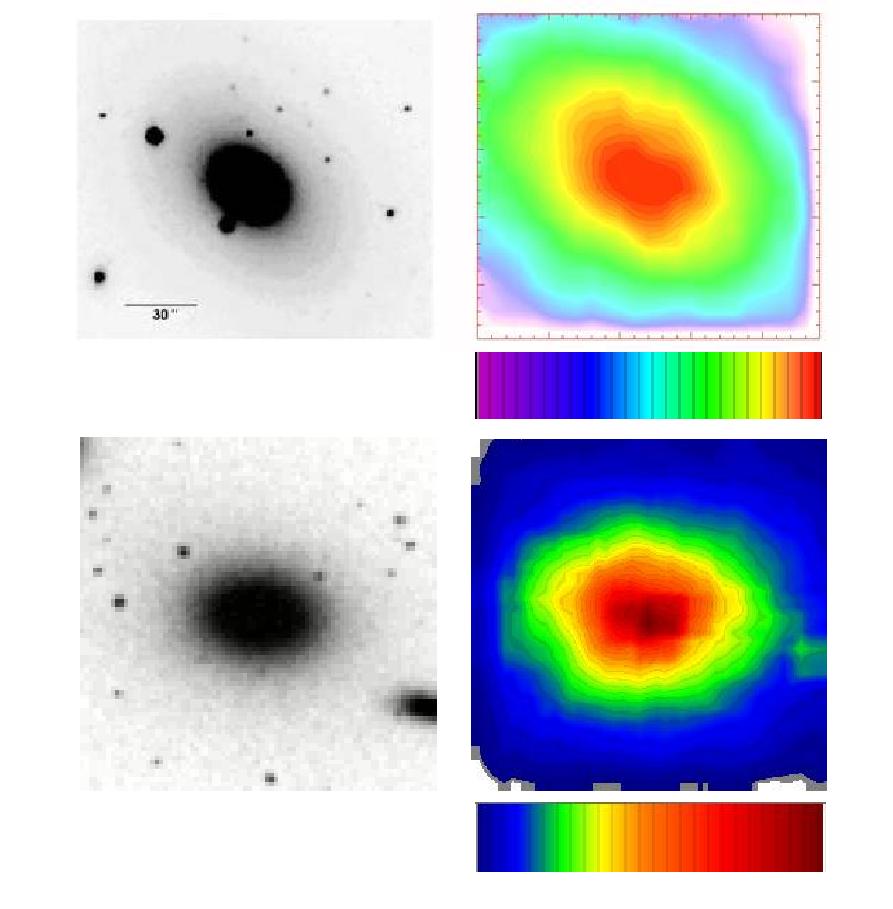}
\caption{\footnotesize{ {\it Left}: 2.5\arcmin\ $\times$2.5\arcmin  R-band DSS image of NGC~4756 (TOP)  and of NGC 5328 (BOTTOM). 
{\it Right}: R map, of YZ projection of the 
selected snapshot, on the same scale as DSS image: 60 equispaced levels with density contrast equal to 
hundred, normalized to the total flux in the map; the lower panel 
shows the color distribution of different levels, from the maximum, 
red, to the minimum value, violet.}}
       \label{confmap}
   \end{figure*}

\section{Modeling}

The optical, X-ray and dynamics observations provide us with a different 
picture of the two groups. NGC 5328 is at the center of its group, while 
NGC 4756 is located to the North-East side.
At about 7\arcmin\ SW from  NGC 4756, a dense agglomerate of galaxies, 
identified  by \citet{GrutzbaN4756} as 
a Hickson-like compact group, is likely evolving as a substructure, contributing $<$ 20\% of the total mass. 
The analysis of the velocities of group members shows that the  velocity dispersions differ by
about a factor of two, larger in NGC~4756, while the harmonic radii are the same.  

Within the limitations imposed by the complexity of the NGC~4756 field, the X-ray  luminosities of NGC 4756 and NGC 5328 are comparable, while the large scale spatial distributions are difficult to compare.  The indication of  emission connecting the two main mass agglomerates in NGC~4756 suggests that hot gas links the two condensations, but we cannot assess its  full scale distribution.

In the above framework,  we consider whether the  formation/evolution of the bright
members, NGC 4756 and NGC 5328, is different or whether they have followed a 
similar path. 
In this latter case, may we identify at what phase of their evolution are we observing them?

To address these questions we attempt to best fit the overall SEDs 
and  global properties of NGC~4756 and NGC~5328, selecting from 
a large set of chemo-photometric SPH simulations the solutions  consistent with the 
dynamical properties of their groups (see below).

Our SPH simulations of galaxy formation and evolution start from the 
same initial conditions described in \cite{MC03} and \cite{Mazzei2003} i.e., collapsing  triaxial systems initially 
composed of dark matter (DM) and gas in  different proportions and 
different total masses. 
We then allowed a large set of galaxy encounters  involving 
systems with a range of mass ratios from 1:1 to 1:10.
In what follows,  minor mergers are excluded since they do not represent the observed global properties as we observe them today. 

In order to exploit a vast range of orbital parameters, we 
carried out different simulations for each pair of interacting 
systems, varying the orbital initial conditions in order to have, 
for the ideal Keplerian orbit of two mass points, the first peri-center 
separation,  p, ranging from the initial length of the major axis 
of the  dark matter triaxial halo of the primary system to  1/10 of the same (major) axis.

For each of these separations, we changed the eccentricity in 
order to have hyperbolic orbits of different energies. 
For the most part we studied
direct encounters, 
where the spins 
of the systems are equal \citep{MC03}, generally parallel to each other, 
and perpendicular to the orbital plane. 
However, we also analyzed some cases with misaligned  spins in 
order to enhance the effects of the system initial rotation on 
the results. Moreover, for a given set of encounters with the 
same orbital parameters  we also examined the role of 
increasing initial gas fractions. These simulations will be 
fully discussed elsewhere (Mazzei, in prep).

All  simulations  include self--gravity of gas, stars and DM, 
radiative cooling, hydrodynamical pressure, shock heating, 
artificial viscosity, star formation (SF) and  feedback from 
evolved stars and type II SNe, and chemical enrichment.
The Initial Mass Function (IMF) is of Salpeter type with 
upper mass limit of 100$\,M_\odot$ and lower mass limit of
0.01$\,M_\odot$ (\citealt{Salpeter}; see \citealt{MC03} and references 
therein for a discussion).

\begin{figure}
{\includegraphics[width=9cm]{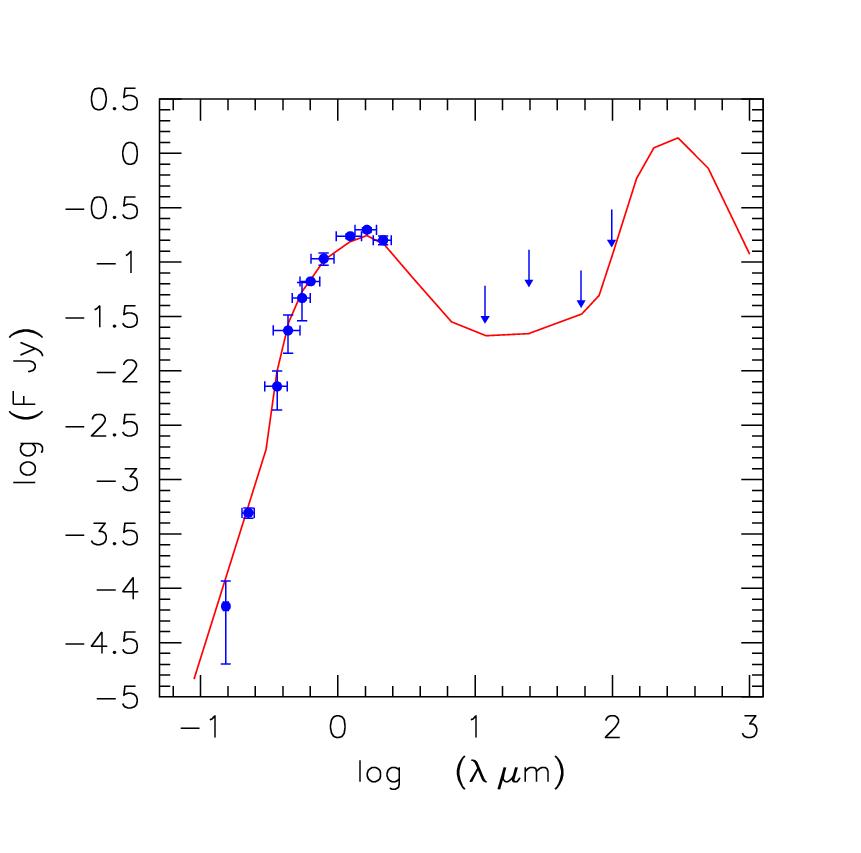}}
\caption{Spectral energy distribution for NGC 4756: (Blue) filled circles correspond to data from  
NED and \citet{GrutzbaN4756} in B and R bands. Arrows show 
upper limits; error bars indicate band width and 3$\sigma$ uncertainties. The continuous line (red) shows the prediction of our model (see
text). 
Dust components (warm and cold with PAH, discussed in \citet{Maeal92}) with  the same average properties derived by  \citet{MaDe94} for a complete  sample of nearby early-type galaxies are included (i.e. I$_0$=46 I$_{local}$, 
Iw=200 I$_{local}$, R$_{wc}$=0.27 and r$_d$=100 r$_c$, see in \citet{Maeal94}).
}
\label{Fig_sedNGC4756}
\end{figure}

\begin{figure}
{\includegraphics[width=\hsize]{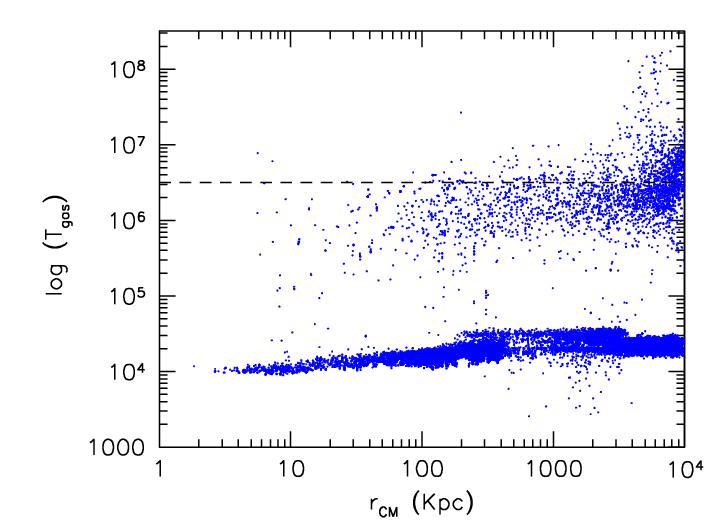}
}
\caption{Radial distribution of the temperature of gas clouds in a frame 
corresponding to the galaxy center of mass weighted by V luminosity.
Each point indicates the simulated gas clouds at the given temperature.
}
\label{gasTemp}
   \end{figure}

As a result, all our simulations provide us with the synthetic SED 
at each evolutionary step, which accounts for chemical 
evolution, stellar  emission, internal extinction and 
re-emission by dust in a self-consistent way, 
as described in \citet[][and references therein]{Spavone09}  and  
extends over almost four orders of magnitude in wavelength,
i.e., from 0.05 to 1000 $\mu$m. So, each simulation self-consistently 
accounts for morphological, dynamical and chemo-photometric evolution. Additionally, we can predict the spatial distribution of clouds of hot gas 
(T$\ge$3$\times$10$^{6}$ K) and total emitted power in the X-ray spectral range, computed following prescriptions of \citet{McKeS79}. 

The whole SEDs  of NGC~5328 and NGC~4756 
and their global properties are well matched by two different snapshots, i.e.,  two different ages,   of the same
simulation. They result from a major merger.
The total initial mass of the systems is  2$\times$10$^{13}$ M$_\odot$ with 
gas  fraction 0.05, so that the total initial mass of the  gas is 
1$\times$10$^{12}$ M$_\odot$. The mass particle resolution is 
5$\times$10$^{7}$ M$_\odot$ for gas and 10$^{9}$ M$_\odot$
for DM particles. This requires 40000 initial particles.

Together with the initial masses, the orbital initial conditions  adopted are: i) the first  peri-center separation is 349~Kpc,  corresponding to 1/5 of the  major axis of the primary system; ii) the orbit eccentricity is 1.3, and  iii) the 
anomaly is 200\degr.
Stars are born in the inner regions of their halos  after about 3 Gyr
from the beginning. Galaxies grow changing their shapes step by 
step as their trajectories are approaching and their halos mixing.

\subsection{NGC 4756}

A passive phase following the turn off of a star formation episode gives a good representation of the global properties of NGC 4756.  The star formation was triggered  7\,Gyr ago  by the strong  interaction between the systems involved in the major merging that produced the galaxy.

Within $\sim$1\,r$_{eff}$(B) [$\sim 12''$, 3.6 kpc] the simulation expects a   stellar population with average age of  
7.5~Gyr, and of 10.5~Gyr inside $\sim$10r$_{eff}$(B) [36 kpc].
The B band  absolute magnitude of the model, -21.06 mag,   agrees well
with the cosmologically corrected value from HyperLeda, -20.83,  and the 
B-R color, 1.76 mag, is consistent with the estimate by \citet{GrutzbaN4756}, 1.65 mag.
Figure \ref{confmap} compares, on the same scale, 2.5\arcmin\ $\times$2.5\arcmin,
the DSS R band image of NGC~4756 with R map of YZ projection of the 
selected snapshot, while  Fig. \ref{Fig_sedNGC4756} compares the predicted SED 
with the available data; the FIR region is not strongly constrained given that only upper limits are available.

\begin{figure}
  \centering
{\includegraphics[width=9cm]{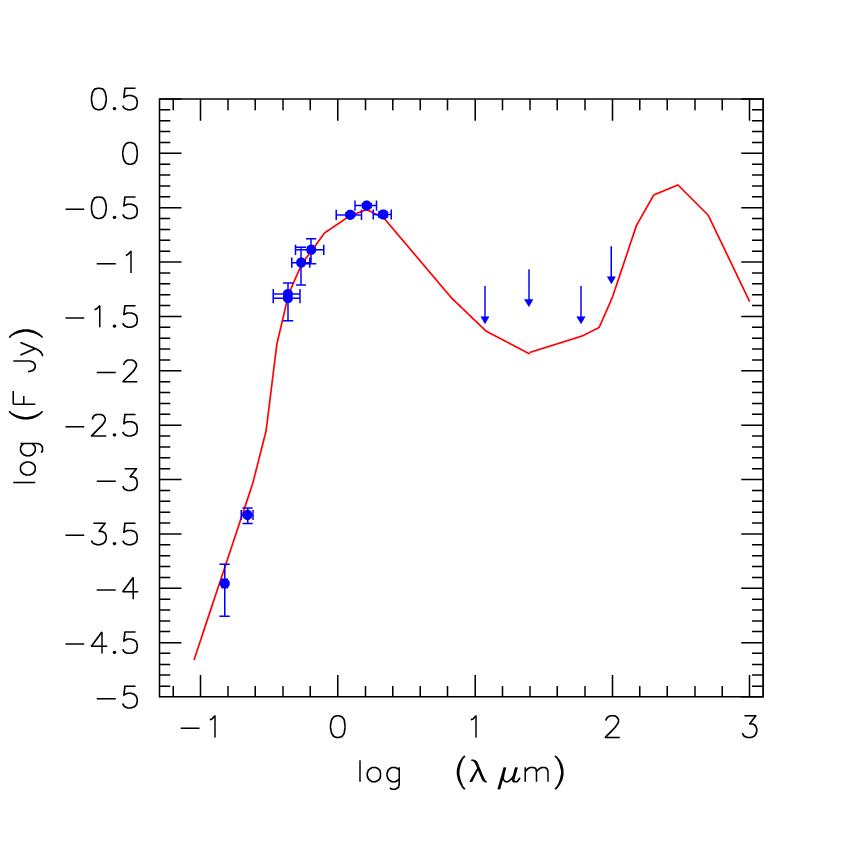}}
\caption{Spectral energy distribution for NGC 5328, as in Fig~\ref{Fig_sedNGC4756}: data from  
NED and \citet{GrutzbaN5328} in B and R bands (Blue filled circles), IR (Arrows for upper limits), model prediction (red continuous line).}
       \label{Fig_sedNGC5328}
   \end{figure}

The SFR of stars  $\le$0.05~Gyr old  is 0.3~M$_\odot$~yr$^{-1}$ and 0.7~M$_\odot$~yr$^{-1}$ inside 2 and 10\,r$_{eff}$(B) respectively.
The rotation velocity, weighed on V luminosity, goes from 
24 km s$^{-1}$ at 4$r_{eff}$(B) up to 40\,km s$^{-1}$ at 15r$_{eff}$(B); the velocity dispersion in the very inner regions 
(r=r$_{eff}$/8$\simeq$0.7\,Kpc) is 174\,km s$^{-1}$  in agreement with
results of $\sim 200$ km s$^{-1}$ by \citet{GrutzbaN4756}. So this system is a slow rotator.

The simulations also provide an estimate of the  mass  and of the X-ray luminosity of the system at the time of the snapshot. 

The total mass expected from our simulation within 157\,Kpc,  which corresponds to the virial radius of the group (where R$_{vir}$=$\pi$/2 $\times$ R$_{H}$=1.57R$_{H}$, with the harmonic radius R$_{H}$ from Table~\ref{din}), is M$\rm _T \sim $
3.73$\times10^{12}\,M_\odot$. Close to 90\% of this mass is in the form of a dark matter halo, since the stellar contribution is $\sim 3 \times 10^{11}\,M_\odot$. 
This value is  well below the virial mass given in Table~\ref{din}, based on the  dynamical analysis of the group.

This can be understood if the simulation is able to explain the formation and evolution of the brightest member for which SED, total stellar luminosity and internal dynamics are well matched by the simulation, but does not reproduce exactly the distribution of the mass, which is on a larger scale than indicated by the dynamical analysis.  The discrepancy in mass could also indicate that the assumption that the group is well virialized is not entirely correct at this scale.  In fact the estimate of the virial radius is relatively unstable, and the inclusion or exclusion of a few members modifies it significantly with smaller variations on the virial mass (see Table~\ref{din}). For example, the simulated and ``measured'' mass within a virial radius of 300 kpc, obtained by ignoring the SW sub-clump in the dynamical analysis,  are in significantly better agreement, 1 vs 2.9 $\times 10^{13}\,M_\odot$.  This is also consistent with the dynamical evidence that the projected mass is larger than the virial mass derived (see \S~\ref{dynanal}).

The spatial distributions of the hot gas 
at the selected snapshot is shown in Fig. \ref{gasTemp}.
In this figure the reference frame  corresponds to the center of mass of the galaxy.
The total luminosity in the X ray spectral range due to gas at T$\ge$3$\times10^6$ K   
is L$_X$=1.7 $\times$10$^{41}$\,erg/s inside 40 kpc, in agreement with the
values measured with XMM-Newton.

\subsection{NGC~5328}

To get a good fit to the global properties of NGC~5328, the simulation needs to be stopped at a time corresponding to an age of  
2.25 Gyr younger  than NGC~4756, so that the last episode of star formation ended about 4.5 Gyr ago.

The average population age inside  1$r_{eff}$(B) [$\sim$53$''$ or 17 kpc]  is 7~Gyr, and 
9~Gyr inside $\sim$ 3$r_{eff}$(B) [$\sim 50$ kpc].  It is interesting to note that 
$r_{eff}$(B) is  about 5 times larger than that of NGC~4756 \citep{GrutzbaN4756}, therefore the regions probed are significantly different in the two objects, which results in similar average ages for the stars.
\cite{Annibali07} provide a luminosity weighted age 
of 12.7$\pm$3.4 of the the nuclear region 
modeling line-strength indices: even considering the large error, 
NGC 5328 is an old object. Weak emission lines from warm gas
are anyway present in the nuclear region of NGC 5328. The perturbation 
induced in the measure of line strength has been taken into account in  \cite{Annibali07} measures. Continuing their study, \cite{Annibali10} 
analyzed the powering mechanism of the warm gas. They found that
the activity class of NGC 5328 nucleus is  ``composite'' i.e.  
it lies in between pure star forming galaxies and LINERs.   
Both observations and our simulation suggest that NGC 5328 is a complex 
object that could have had at intermediate age an episode of star formation.

The rotation velocity, weighed on V luminosity, 
ranges from 22\,km s$^{-1}$  to 40\,km s$^{-1}$  
[1$r_{eff}$(B) -- 3$r_{eff}$(B)].
Moreover, the velocity dispersion inside 2\,Kpc,
corresponding to r$_{eff}$(B)/8, is 276\,km s$^{-1}$, 
in  agreement with the observations \citep{GrutzbaN5328}.
So, according to our results,  this galaxy is  a slow rotator like NGC~4756.
The B band absolute magnitude of the model, -21.22 mag, is consistent
with the cosmologically corrected value from HyperLeda, 
-21.76,  and the B-R color, 1.80 mag,  with the value of  1.79 mag by 
\citet{GrutzbaN5328}.

The total mass expected within 1\,$r_{eff}$(B)
is 3.75$\times10^{11}\,M_\odot$, with  $\sim$26$\%$ of dark matter.
Figure~\ref{confmap} compares the DSS R band image of NGC~5328 
with R map of YZ projection of the selected snapshot, and 
Figure~\ref{Fig_sedNGC5328} the predicted SED 
with the available data (see Fig. \ref{Fig_sedNGC4756} for details).

Predictions from this snapshot are consistent with the dynamical analysis 
of the group itself (\S~\ref{dynanal} and  Table~\ref{din}). The total mass expected within 157\,kpc, i.e., the virial radius of the group, 3.73$\times10^{12}\,M_\odot$,
is in agreement with  the virial mass of the group itself.
Similarly, the total luminosity is  L$_X$=2.91
$\times$10$^{41}$\,erg/s  inside 110\,kpc, 
consistent with the values measured from the XMM-Newton data.

\section{Summary and Conclusions}

We have detected both galaxies with XMM-Newton data, and measured a sizable 
contribution from hot gas in both systems. The extent of the emission is larger 
than the optical size of the galaxies. We measure a total size of $\sim$110 kpc 
for NGC 5328. NGC 4756 is observed against an X-ray emitting background cluster 
which prevents us from properly measuring the total extent of the source.  However 
we detect emission out to a radius of $\sim$40 kpc around NGC 4756, and a 
connection to the SW compact-like group at $\sim $130 kpc that is presumably 
interacting with NGC 4756 itself. The total luminosities measured are L$_x \sim 10^{41}$ \ergs, relatively low for groups, but this should be considered a lower limit for NGC 4756. 

The results of SPH simulations, which describe the dynamical, morphological and chemo-photometric evolution in a self-consistent way,  suggests that NGC 4756 and NGC 5328 can be described by the same evolutionary path, stopped at two different epochs, with the last episode of star formations at 7 and 4.5 Gyr ago, respectively. The data and simulations sketch  
old and digested merging events at their origin, and they can now both be considered mature systems.  The predictions are in good agreement with observed quantities such as SED, $\sigma$ for stars, and X-ray luminosity. 

In spite of the apparent similarity of these two galaxies, in terms of luminosity, internal rotation, stellar content and evolutionary path, the evidence suggests that their respective groups are at different evolutionary phases.

A new analysis of the dynamics of both systems 
confirms that NGC 5328 sits at the center of a small group, which appears to be 
azimuthally symmetric and  relatively relaxed. NGC 5328 is the dominant galaxy in the system, about 2 mag. brighter than the next bright object.  The group has a relatively low virial mass, of a few $\sim 10^{12}$ M$_\odot$,  comparable to what is expected from a bright early-type galaxy.  The presence of  Abell 3574  at the same redshift suggests that this could be a satellite galaxy at the very periphery of the much larger potential, that has gone through a major preprocessing before being [eventually] acquired by the cluster. 

NGC 4756 is composed of two main 
sub-condensations, NGC 4756 itself and the compact-like small group. The two are at the same average redshift, but might 
not have reached a dynamical equilibrium yet. The evidence of a connection in the X-ray emission between the two condensations suggests an interaction between the two structures which might eventually merge at a later stage of the 
evolution of the group.

\begin{acknowledgements}
We wish to thank the EPIC blank sky team, Jenny Carter and Andy Read, for the effort they put in helping us solve our background issues, and the clever suggestions they have  made. We thank Giovanni Fasano for 
having kindly provided us with a V-band image of Abell 1361 observed at ESO-MPI2.2m 
with WFI as part of the WINGS project.
We acknowledge a financial contribution from the agreement ASI-INAF
      I/009/10/0.  This research has made
use of the SAOImage DS9, developed by Smithsonian Astrophysical
Observatory and of the NASA/IPAC Extragalactic Database (NED) which
is operated by the Jet Propulsion Laboratory, California Institute of
Technology, under contract with the National Aeronautics and Space
Administration. {\tt IRAF} is distributed
by the National Optical Astronomy Observatories, which are operated
by the Association of Universities for Research in Astronomy, Inc.,
under cooperative agreement with the National Science Foundation.
The Digitized Sky Survey (DSS) was produced at the
Space Telescope Science Institute under U.S. Government grant NAG
W-2166. The images of these surveys are based on photographic data
obtained using the Oschin Schmidt Telescope at the Palomar
Observatory and the UK Schmidt Telescope. The plates were processed
into the present compressed digital form with the permission of
these institutions.
We acknowledge the usage of the HyperLeda database (http://leda.univ-lyon1.fr).
\end{acknowledgements}

\begin{appendix}
\onecolumn
\section{Detected sources in the fields of NGC 4756 and NGC 5328}
\label{appendix}

Besides the two galaxies themselves, the detection algorithm has detected 68 and 32 sources in the fields of NGC 4756 and  NGC 5328, respectively. For this latter, we have excluded a region of 1$'$ radius around NGC 5328, where sources found by detect most likely represent just small fluctuations in the extended emission.  We used a broad energy band 0.5-4.5 keV and a single conversion factor to convert the detected counts in flux, corresponding to a power law model with $\Gamma=1.7$ and N$_H$=3$\times 10^{20}$ cm$^{-2}$ (see  2XMM
catalog). In Table~\ref{n4756-tab} and \ref{n5328-tab} we give the full list of detected 
sources, excluding NGC 4756 and  NGC 5328 themselves (see Table~\ref{members}).  We give the counts and average flux derived from the three instruments combined, when available.  For sources detected in only a subset of the instruments, we give the counts and flux from the specific instrument, as noted in the tables. Identifications with optical objects is given when available, or a comment when  the positional coincidence indicates a clear association.

\begin{longtable}{lllrrcl}
\caption{Detected sources in the field of NGC 4756.  
\label{n4756-tab}}
\\
\hline 
\hline
Nr.  & $\alpha$ &$\delta$& Counts& Error& $f_x$ (0.5-4.5 keV) &identification\\
&(J2000.0) &  (J2000.0) &\multicolumn{2}{c}{0.5-4.5 keV } &\ergcms \\
\hline
\endfirsthead
\caption{continued.}\\
\hline
\hline
{Nr.}   & {$\alpha$} & {$\delta$}& Counts& Error& $f_x$ (0.5-4.5 keV)&identification\\
&(J2000.0)  & (J2000.0)  &\multicolumn{2}{c}{0.5-4.5 keV } &\ergcms \\
\hline
\endhead
\hline
\endfoot
\hline
1&12:51:49.623  &-15:32:21.25&  271.3  &  20.4 & 3.3$\times 10^{-14}$ & Background gal. \\
2&12:51:52.791  &-15:29:36.60&  119.4  &  16.0 & 1.3$\times 10^{-14}$ & \\ 
3&12:51:53.713  &-15:24:47.18&   65.9  &  12.5 & 7.9$\times 10^{-15}$ &faint object \\
4&12:51:54.418  &-15:33:10.47&   85.5  &  14.0 & 8.8$\times 10^{-15}$  \\
5$^\dagger$&12:51:55.915  &-15:18:48.28&   38.5  &  9.2 &  1.1$\times 10^{-14}$  \\
6&12:52:01.404  &-15:34:29.50&  342.6  &  22.5 & 3.4$\times 10^{-14}$ &\\ 
7&12:52:01.580  &-15:22:28.62&   96.8  &  14.6 & 8.7$\times 10^{-15}$  \\
8&12:52:03.609  &-15:39:31.34&   49.9  &  11.2 & 2.7$\times 10^{-15}$ & Point-like source?  \\
9&12:52:03.881  &-15:39:04.38&  129.8  &  16.4 & 1.6$\times 10^{-14}$  \\
10&12:52:04.073  &-15:27:33.42&   54.3  &  13.2 & 4.1$\times 10^{-15}$ & next to bright!\\
11&12:52:04.147  &-15:35:58.62&  108.0  &  15.2 & 1.0$\times 10^{-14}$  \\
12&12:52:04.161  &-15:27:09.39&   99.5  &  15.6 & 7.5$\times 10^{-15}$  \\
13&12:52:04.622  &-15:23:57.42&  368.2  &  23.8 & 3.0$\times 10^{-14}$ &faint object\\
14&12:52:06.664  &-15:35:54.55&   73.9  &  13.8 & 6.3$\times 10^{-15}$ \\
15&12:52:07.715  &-15:26:17.79&  445.9  &  26.1 & 3.3$\times 10^{-14}$ & faint object\\
16&12:52:10.984  &-15:20:16.18&   49.7  &  12.1 & 3.8$\times 10^{-15}$   \\
17&12:52:11.796  &-15:24:38.70&  156.5  &  17.9 & 1.0$\times 10^{-14}$  \\
18&12:52:11.822  &-15:19:11.73&  228.6  &  25.1 & 2.1$\times 10^{-14}$  \\
19&12:52:14.795  &-15:34:23.98&  192.1  &  19.0 & 1.5$\times 10^{-14}$ & USGC S188\\
20&12:52:15.048  &-15:23:47.64&   76.8  &  13.7 & 5.1$\times 10^{-15}$ &\\ 
21&12:52:15.998  &-15:17:05.75&   76.9  &  14.7 & 8.5$\times 10^{-15}$  \\
22&12:52:18.912  &-15:21:47.07&  122.1  &  19.5 & 1.2$\times 10^{-14}$ & at the border of a bright object \\
23&12:52:19.527  &-15:41:24.46&  222.0  &  40.9 & 2.8$\times 10^{-14}$   \\
24&12:52:23.679  &-15:22:50.09&  112.0  &  15.8 & 6.3$\times 10^{-15}$ & bright Point-like Source \\
25&12:52:23.760  &-15:30:12.39&  142.6  &  18.0 & 9.4$\times 10^{-15}$ &\\ 
26&12:52:24.129  &-15:33:27.32&   83.6  &  14.4 & 5.2$\times 10^{-15}$  \\
27&12:52:25.149  &-15:32:16.74&  136.6  &  17.2 & 7.4$\times 10^{-15}$ & Background gal. z=0.151461 \\
28&12:52:27.148  &-15:31:05.81&  470.4  &  43.1 & 2.6$\times 10^{-14}$ & IC 0829\\
29&12:52:28.301  &-15:32:24.75&   73.1  &  17.0 & 3.4$\times 10^{-15}$   \\
30&12:52:29.912  &-15:34:32.16&  131.9  &  18.3 & 9.1$\times 10^{-15}$ &\\ 
31&12:52:31.041  &-15:30:37.22&   61.0  &  15.0 & 2.4$\times 10^{-15}$  \\
32$^\dagger$&12:52:31.334  &-15:17:18.18&  169.9 &  16.3 &3.4$\times 10^{-14}$&faint object \\
33&12:52:31.424  &-15:40:32.81&   89.9  &  14.8 & 9.1$\times 10^{-15}$  \\
34&12:52:32.113  &-15:39:43.08&   56.5  &  13.2 & 3.8$\times 10^{-15}$ & Point-like source?  \\
35&12:52:32.157  &-15:28:52.90&  266.2  &  28.7 & 1.3$\times 10^{-14}$ \\
36&12:52:32.405  &-15:28:41.90&  997.9  &  42.8 & 5.4$\times 10^{-14}$ & Background  point-like source? \\ 
37&12:52:33.154  &-15:30:59.84& 2111.0  &  52.2 & 1.0$\times 10^{-13}$ & MCG -02-33-038\\
38$^\dagger$&12:52:36.126  &-15:18:20.10&  78.1& 12.5 & 1.4$\times 10^{-14}$& Gal in A1361 \\
39&12:52:37.950  &-15:31:51.01&  237.8  &  21.3 & 1.1$\times 10^{-14}$ &faint object \\
40&12:52:39.329  &-15:34:57.21&   43.3  &  12.5 & 3.3$\times 10^{-15}$  & Point-like source \\
41&12:52:40.258  &-15:41:22.77&   75.8  &  15.5 & 5.6$\times 10^{-15}$   \\
42&12:52:40.777  &-15:32:04.98&   59.9  &  13.4 & 2.6$\times 10^{-15}$  & Gal in A1361 \\
43&12:52:41.277  &-15:26:04.90&  443.1  &  27.7 & 1.9$\times 10^{-14}$  \\
44&12:52:43.931  &-15:35:20.65&  106.8  &  18.0 & 7.1$\times 10^{-15}$  \\
45&12:52:44.296  &-15:40:25.71&   42.0  &  11.0 & 1.0$\times 10^{-15}$ & Point-like source?  \\
46&12:52:45.037  &-15:28:32.55&  239.5  &  22.0 & 1.0$\times 10^{-14}$  \\
47&12:52:45.619  &-15:25:19.94&  417.0  &  27.9 & 1.9$\times 10^{-14}$  &Background  point-like source?\\
48&12:52:48.390  &-15:26:26.97&  491.8  &  30.0 & 2.2$\times 10^{-14}$ \\
49$^\dagger$&12:52:50.291  &-15:22:15.60&  71.8&  13.2&9.1$\times 10^{-15}$  \\
50$^\dagger$&12:52:50.385  &-15:19:38.71&  155.2&  16.5 & 2.5$\times 10^{-14}$&faint object \\
51&12:52:50.469  &-15:24:04.39&  915.6  & 104.1 & 5.0$\times 10^{-14}$ & Point-like source in halo of NGC 4756\\
52&12:52:52.749  &-15:35:44.31&  135.0  &  17.7 & 9.2$\times 10^{-15}$  \\
53&12:52:54.319  &-15:41:31.59&  780.6  &  85.9 & 8.1$\times 10^{-15}$  \\
54$^\dagger$&12:52:56.058  &-15:20:30.42&   86.1&  14.5 & 1.4$\times 10^{-14}$& Gal in A1361 \\
55$^\dagger$&12:52:57.588  &-15:20:44.61&   56.7 &  12.6 & 8.9$\times 10^{-15}$ & Gal in A1361 \\
56&12:52:58.104  &-15:34:51.93&  104.9  &  16.9 & 8.1$\times 10^{-15}$  \\
57&12:53:01.051  &-15:39:38.33&  472.4  &  27.3 & 4.8$\times 10^{-14}$ & faint object \\
58$^\dagger$&12:53:02.525  &-15:21:22.32&    56.3  &   11.7 & 9.5$\times 10^{-15}$  \\
59&12:53:05.937  &-15:39:11.63&  102.0  &  16.3 & 1.0$\times 10^{-14}$  \\
60&12:53:08.930  &-15:34:21.28&   92.5  &  15.1 & 6.4$\times 10^{-15}$  \\
61&12:53:10.535  &-15:26:24.85&  179.7  &  18.9 & 1.2$\times 10^{-14}$ \\
62$^\dagger$&12:53:11.936  &-15:21:16.02&  40.5 &  10.7  & 7.4$\times 10^{-15}$   \\
63&12:53:18.411  &-15:32:04.01& 2226.1 &  63.7 & 1.8$\times 10^{-13}$ & Gal in A1361\\
64&12:53:20.118  &-15:25:40.59&  106.2  &  16.7 & 8.2$\times 10^{-15}$ &bright object \\
65&12:53:25.348  &-15:28:09.22&  270.4  &  22.0 & 2.6$\times 10^{-14}$  & Background gal.\\
66$^{\dagger\dagger}$&12:53:26.011  &-15:33:30.90&  45.2 &  10.3 &1.6$\times 10^{-14}$    \\
67$^{\dagger\dagger\dagger}$&12:53:30.483  &-15:22:31.08&    32.3 &   8.7 & 1.2$\times 10^{-14 }$ \\
68$^{\dagger\dagger\dagger}$&12:53:32.550  &-15:32:03.40&   32.8 &   9.0 & 1.2$\times 10^{-14}$  \\

\hline
\end{longtable}
\noindent $^\dagger$ EPIC-pn data.\\
$^{\dagger\dagger}$  EPIC-M1 data\\
$^{\dagger\dagger\dagger}$ EPIC-M2 data

\begin{longtable}{lllrrcl}
\caption{Detected sources in the field of NGC5328. 
\label{n5328-tab}}
\\
\hline 
\hline
Nr.  & $\alpha$ &$\delta$& Counts& Error& $f_x$ (0.5-4.5 keV) &identification\\
&(J2000.0) &  (J2000.0) &\multicolumn{2}{c}{0.5-4.5 keV } &\ergcms \\
\hline
\endfirsthead
\caption{continued.}\\
\hline
\hline
{Nr.}   & {$\alpha$} & {$\delta$}& Counts& Error& $f_x$ (0.5-4.5 keV)&identification \\
&(J2000.0)  & (J2000.0)  &\multicolumn{2}{c}{0.5-4.5 keV } &\ergcms \\
\hline
\endhead
\hline
\endfoot
\hline

1$^\dagger$&13:51:52.001& -28:27:27.35  &   87.7 &  20.2  & 5.0$\times 10^{-14}$  &  2MASXJ13515205-2827344 z=0.037336 \\
2&13:52:00.238 &-28:25:34.53  &   53.9 &  13.0  & 1.8$\times 10^{-14}$  &    \\
3&13:52:10.571 &-28:25:05.76  &   20.6 &   7.9  & 1.9$\times 10^{-15}$ &    \\
4&13:52:16.906 &-28:30:41.24  &   24.1 &   8.3  & 2.3$\times 10^{-15}$   &    \\
5&13:52:18.030 &-28:27:46.10 &    85.2&   14.6 &  1.9$\times 10^{-14}$ &     \\
6&13:52:20.045 &-28:38:16.72 &    63.3&   13.2 &  1.9$\times 10^{-14}$ &     \\
7&13:52:20.886 &-28:33:59.91 &    46.2&   12.1 &  8.3$\times 10^{-15}$ &     \\
8$^\dagger$&13:52:22.080 &-28:31:17.68 &    86.4&   15.3 &  2.5$\times 10^{-14}$  &    faint object  \\
9&13:52:26.736 &-28:22:22.45 &   148.2&   17.1 &  3.1$\times 10^{-14}$ &   faint object   \\
10&13:52:26.870 &-28:31:53.59 &   265.8&   21.3 &  4.8$\times 10^{-14}$ &  point source    \\
11$^{\dagger\dagger}$&13:52:27.550 &-28:41:53.07 &    30.1&    7.3 &  3.6$\times 10^{-14}$  & bright object (star?)    \\
12&13:52:33.988 &-28:28:50.22 &    47.1&   12.2 &  4.7$\times 10^{-15}$ &    \\
13&13:52:38.310 &-28:15:58.81 &    45.1&   12.3 &  1.3$\times 10^{-14}$ &  star  \\
14&13:52:39.980 &-28:18:03.54 &    77.3&   14.2 &  2.0$\times 10^{-14}$ &    \\
15&13:52:44.040 &-28:23:56.25 &   200.2&   19.0 &  3.0$\times 10^{-14}$  &    \\
16&13:52:48.781 &-28:29:59.46 &   517.1&  115.7 &  6.7$\times 10^{-14}$  &   PGC3094716  \\
17&13:52:49.010 &-28:35:31.03 &    84.3&   13.9 &  1.4$\times 10^{-14}$&    \\
18&13:52:49.283 &-28:28:10.60 &    60.2&   15.0 &  6.0$\times 10^{-15}$  &    \\
19$^\dagger$&13:52:50.496 &-28:17:47.37 &    52.6&   11.9 &  1.8$\times 10^{-14}$ &    \\
20&13:52:53.884 &-28:31:46.93 &    70.1&   14.7 &  8.0$\times 10^{-15}$  &   2MASXJ13525393-2831421 \\
21&13:52:54.091 &-28:30:35.46 &   604.1&   52.5 &  9.1$\times 10^{-14}$  &    \\
22&13:52:57.238 &-28:28:01.18 &    63.8&   15.3 &  7.0$\times 10^{-15}$  &    \\
23&13:52:59.138 &-28:28:21.73 &   221.4&   38.4 &  2.5$\times 10^{-14}$  &   NGC5330\\
24&13:52:59.887 &-28:35:23.05 &   150.4&   17.2 &  2.4$\times 10^{-14}$  &   CXO J135259.8-283523,  bright object \\
25&13:53:02.056 &-28:40:44.45 &   110.1&   15.4 &  2.8$\times 10^{-14}$  &    \\
26$^{\dagger\dagger\dagger}$&13:53:03.219 &-28:14:52.84 &   103.8&   11.6 &  2.1$\times 10^{-13}$ &  CXO J135303.0-281502 \\
27&13:53:08.095 &-28:34:05.41 &    93.6&   14.8 &  1.1$\times 10^{-14}$  &    \\
28&13:53:14.928 &-28:25:40.86 &    49.6&   13.6 &  1.2$\times 10^{-14}$  &   ESO445-070 \\
29$^\dagger$&13:53:27.950 &-28:23:02.20 &   303.3&   21.6 &  7.6$\times 10^{-14}$  &   faint object\\
30&13:53:31.424 &-28:34:07.20 &    47.1&   11.8 &  9.2$\times 10^{-15}$ &    \\
31$^{\dagger\dagger}$&13:53:40.261 &-28:35:55.30 &    34.7&    8.1 &  2.4$\times 10^{-14}$ &   \\
32$^{\dagger\dagger}$&13:53:52.645 &-28:29:07.54 &    25.4&    7.0 &  2.3$\times 10^{-14}$ &  \\

\hline
\end{longtable}
\noindent $^\dagger$  EPIC-pn data \\
$^{\dagger\dagger}$ EPIC-M1 + EPIC-M2 data.\\
$^{\dagger\dagger\dagger}$ EPIC-M2 data.
\end{appendix}
\end{document}